\documentclass[pra,aps,showpacs,epsf,floats,10pt]{revtex4}

\usepackage{amssymb}
\usepackage{amsfonts}
\usepackage{amsmath}

\begin{document}

\title{A Geometric Algebra Perspective On Quantum Computational Gates And
Universality In Quantum Computing}
\author{Carlo Cafaro}
\email{carlo.cafaro@unicam.it}
\affiliation{Dipartimento di Fisica, Universit\`{a} di Camerino, I-62032 Camerino, Italy}
\author{Stefano Mancini}
\email{stefano.mancini@unicam.it}
\affiliation{Dipartimento di Fisica, Universit\`{a} di Camerino, I-62032 Camerino, Italy}

\begin{abstract}
We investigate the utility of geometric (Clifford) algebras (GA) methods in
two specific applications to quantum information science. First, using the
multiparticle spacetime algebra (MSTA, the geometric algebra of a
relativistic configuration space), we present an \emph{explicit} algebraic
description of one and two-qubit quantum states together with a MSTA
characterization of one and two-qubit quantum computational gates. Second,
using the above mentioned characterization and the GA description of the Lie
algebras $SO\left( 3\right) $ and $SU\left( 2\right) $ based on the rotor
group $\emph{Spin}^{+}\left( 3\text{, }0\right) $ formalism, we reexamine
Boykin's proof of universality of quantum gates. We conclude that the MSTA
approach does lead to a useful conceptual unification where the \emph{complex%
} qubit space and the complex space of unitary operators acting on them
become united, with both being made just by multivectors in \emph{real}
space. Finally, the GA approach to rotations based on the rotor group does
bring conceptual and computational advantages compared to standard vectorial
and matricial approaches.
\end{abstract}

\maketitle

\section{Introduction}

Geometric (Clifford) algebra (GA) \cite{hestenes, dl} is a universal
language for physics based on the mathematics of Clifford algebra.
Applications of GA in physics span from quantum theory and gravity \cite%
{gull, francis} to classical electrodynamics \cite{baylis} reaching even
massive classical electrodynamics with Dirac's magnetic monopoles \cite%
{cafaro, cafaro-ali}.

There are some natural ways of including Clifford algebra and GA in quantum
information science (QIS) motivated by physical reasons \cite{vlasov, doran1}%
. For instance, any qubit (quantum bit, elementary carrier of quantum
information) modelled as a spin-$\frac{1}{2}$ system can be regarded as a $%
2\times 2$ matrix with an empty second column. All $2\times 2$ matrices are
combinations of Pauli matrices which are a representation of some GA. All $%
2\times 2$ unitary transformations can be parametrized by elements of GA as
well. Driven by such motivations, the first \emph{formal} reformulations of
some of the most important operations of quantum computing in the
multiparticle geometric algebra formalism have been presented \cite{somaroo}%
. In a less conventional and very recent GA approach to quantum computing,
the possibility of performing quantum-like algorithms using GA structures
without involving quantum mechanics has been explored \cite{marek1, marek2,
marek3}. Within this approach, the standard tensor product is replaced by
the geometric product and entangled states are replaced by multivectors with
a geometrical interpretation in terms of "\emph{bags of shapes}". Such GA
approach brings new conceptual elements in QIS and the formalism of quantum
computation loses its microscopic flavor when viewed from this novel GA
point of view. Indeed, non-microworld implementations of quantum computing
are suggested since there is no fundamental reason to believe that quantum
computation has to be associated with physical systems described by quantum
mechanics \cite{marek3}.

In \cite{doran1}, an extended discussion on applications of GA\ techniques
in quantum information is presented, however the fundamental concept of
universality in quantum computing is not discussed. In \cite{somaroo}, the
GA formulation of the Toffoli and Fredkin three-qubit quantum gates is
introduced but no explicit characterization of all one and two-qubit quantum
gates appears. Here, inspired by these two works, we present a
(complementary and self-contained) compact, explicit and expository GA
characterization of one and two qubit quantum gates together with a novel
GA-based perspective on the concept of universality in quantum computation.%
\textbf{\ }First, we present an \emph{explicit} multiparticle spacetime
algebra (MSTA, the geometric Clifford algebra of a relativistic
configuration space) \cite{lasenby93, cd93, doran96, somaroo99} algebraic
description of one and two-qubit quantum states (for instance, the $2$-qubit
Bell states) together with a MSTA characterization of one (bit-flip,
phase-flip, combined bit and phase flip quantum gates, Hadamard gate,
rotation gate, phase gate and $\frac{\pi }{8}$-gate) and two-qubit quantum
computational gates (CNOT, controlled-phase and SWAP quantum gates) \cite%
{NIELSEN}. Second, using the above mentioned explicit characterization and
the GA description of the Lie algebras $SO\left( 3\right) $ and $SU\left(
2\right) $ based on the rotor group $\emph{Spin}^{+}\left( 3\text{, }%
0\right) $ formalism, we reexamine Boykin's proof of universality of quantum
gates \cite{B99, B00}. We conclude that the MSTA approach does lead to a
useful conceptual unification where the complex qubit space and the complex
space of unitary operators acting on them become united, with both being
made just by multivectors in real space. Furthermore, the GA approach to
rotations based on the rotor group clearly brings conceptual and
computational advantages compared to standard vectorial and matricial
approaches.

The layout of this article is as follows. In Section II, the basic MSTA
formalism for the GA characterization of elementary gates for quantum
computation is presented. In Section III, we present an explicit GA
characterization of one and two-qubit quantum states together with a GA
characterization of one and two-qubit quantum computational gates.
Furthermore, we briefly discuss the extension of the MSTA formalism to
density matrices. \ In Section IV, using the above mentioned explicit
characterization and the GA description of the Lie algebras $SO\left(
3\right) $ and $SU\left( 2\right) $ based on the rotor group $\emph{Spin}%
^{+}\left( 3\text{, }0\right) $ formalism, we reexamine Boykin's proof of
universality of quantum gates \cite{B99, B00}. Finally, our final remarks
are presented in Section V.

\section{Multiparticle Spacetime Algebra}

In this Section, we present the basic MSTA formalism for the GA
characterization of elementary gates for quantum computation.

\subsection{The $n$-Qubit Spacetime Algebra Formalism}

It is commonly believed that complex space notions and an imaginary unit $i_{%
\mathbb{C}
}$ are fundamental in quantum mechanics. However using spacetime algebra
(STA, the geometric Clifford algebra of real $4$-dimensional Minkowski
spacetime, \cite{dl}) it has been shown how the $i_{%
\mathbb{C}
}$ appearing in the Dirac, Pauli and Schrodinger equations has a geometrical
interpretation in terms of rotations in real spacetime \cite{david75}. This
becomes clear once introduced the geometric algebra of a relativistic
configuration space, the so-called multiparticle spacetime algebra (MSTA) 
\cite{lasenby93, cd93, doran96, somaroo99}.

In the orthodox formulation of quantum mechanics, the tensor product is used
in constructing both multiparticle states and many of the operators acting
on these states. It is a notational device for explicitly isolating the
Hilbert spaces of different particles. Geometric algebra attempts to justify
from a foundational point of view the use of the tensor product in
nonrelativistic quantum mechanics in terms of the underlying geometry of
space-time \cite{cd93}. The GA formalism provides an alternative
representation of the tensor product in terms of the\textbf{\ }\emph{%
geometric product}\textbf{.} Motivated by the usefulness of the STA\
formalism in describing a single-particle quantum mechanics, the MSTA\
approach to multiparticle quantum mechanics in both non-relativistic and
relativistic settings was originally \cite{cd93} introduced with the hope
that it would also provide both computational and, above all,
interpretational advances in multiparticle quantum theory. Conceptual
advances are expected to arise by exploiting the special geometric insights
that the MSTA approach provides. The unique feature of the MSTA\ is that it
implies a separate copy of the time dimension for each particle, as well as
the three spatial dimensions. It constitutes an attempt to construct a solid
conceptual framework for a multi-time approach to quantum theory. Therefore,
the main original motivation for using such formalism is the possibility of
shedding light on issues of\textbf{\ }\emph{locality} and \emph{causality}
in quantum theory. Indeed, interesting applications of the MSTA method
devoted to the reexamination of Holland's causal interpretation of a system
of two spin-$\frac{1}{2}$\ particles \cite{holland} appear in \cite{doran96,
somaroo99}. Following this line of investigation, in this article we apply
the MSTA method to express qubits and quantum gates and to revisit in
geometric algebra terms Boykin's proof of universality in quantum computing.

The multiparticle spacetime algebra provides the ideal algebraic structure
for studying multiparticle states and operators. MSTA is the geometric
algebra of $n$-particle configuration space which, for relativistic systems,
consists of $n$ copies (each copy is a $1$-particle space) of Minkowski
spacetime. A suitable basis for the MSTA is given by the set $\left\{ \gamma
_{\mu }^{a}\right\} $, where $\mu =0$,.., $3$ labels the spacetime vector
and $a=1$,.., $n$ labels the particle space. These basis vectors satisfy the
orthogonality conditions $\gamma _{\mu }^{a}\cdot \gamma _{\nu }^{b}=\delta
^{ab}\eta _{\mu \nu }$ with $\eta _{\mu \nu }=$ diag$\left( +\text{, }-\text{%
, }-\text{, }-\right) $. Vectors from different particle spaces anticommute
as a consequence of their orthogonality. Note that a basis for the entire
MSTA has $2^{4n}$ degrees of freedom, $\dim _{%
\mathbb{R}
}\left[ \mathfrak{cl}\left( 1\text{, }3\right) \right] ^{n}=2^{4n}$. In
nonrelativistic quantum mechanics, all of the individual time coordinates
are identified with a single absolute time. We take this vector to be $%
\gamma _{0}^{a}$ for each $a$. Spatial vectors relative to these timelike
vectors are modeled as bivectors through a spacetime split. A basis set of
relative vectors is then defined by $\sigma _{k}^{a}\overset{\text{def}}{=}%
\gamma _{k}^{a}\gamma _{0}^{a}$, with $k=1$,.., $3$ and $a=1$,.., $n$. For
each particle space the set $\left\{ \sigma _{k}^{a}\right\} $ generates the
geometric algebra of relative space $\mathfrak{cl}(3)\cong \mathfrak{cl}%
^{+}(1$, $3)$. Each particle space has a basis given by,%
\begin{equation}
1\text{, }\left\{ \sigma _{k}\right\} \text{, }\left\{ i\sigma _{k}\right\} 
\text{, }i\text{,}  \label{bf}
\end{equation}%
where the volume element $i$ (the \textit{pseudoscalar}, the highest grade
multivector) is defined by $i\overset{\text{def}}{=}\sigma _{1}\sigma
_{2}\sigma _{3}$ (suppressing the particle space indices). The basis in (\ref%
{bf}) defines the Pauli algebra (the geometric algebra of the $3$%
-dimensional Euclidean space, \cite{dl}) but in GA\ the three Pauli $\sigma
_{k}$ are no longer viewed as three matrix-valued components of a single
isospace vector, but as three independent basis vectors for real space.
Notice that unlike spacetime basis vectors, relative vectors $\left\{ \sigma
_{k}^{a}\right\} $ from separate particle spaces commute, $\sigma
_{k}^{a}\sigma _{j}^{b}=\sigma _{j}^{b}\sigma _{k}^{a}$, $a\neq b$. It turns
out that the $\left\{ \sigma _{k}^{a}\right\} $ generate the direct product
space $\left[ \mathfrak{cl}(3)\right] ^{n}\overset{\text{def}}{=}\mathfrak{cl%
}(3)\otimes $...$\otimes \mathfrak{cl}(3)$ of $n$ copies of the geometric
algebra of the $3$-dimensional Euclidean space. Within the MSTA\ formalism,
Pauli spinors (a spin-$\frac{1}{2}$ quantum system is an adequate model of
quantum bit) may be represented as elements of the even subalgebra of the
Pauli algebra spanned by $\left\{ 1\text{, }i\sigma _{k}\right\} $ and
isomorphic to the quaternion algebra. This space is a $4$-dimensional real
space where a general even element can be written as, $\psi
=a^{0}+a^{k}i\sigma _{k}$, where $a^{0}$ and $a^{k}$ with $k=1$, $2$, $3$
are \textit{real} scalars. An ordinary quantum state contains a pair of 
\textit{complex }numbers, $\alpha $ and $\beta $,%
\begin{equation}
\left\vert \psi \right\rangle =\left( 
\begin{array}{c}
\alpha \\ 
\beta%
\end{array}%
\right) =\left( 
\begin{array}{c}
\mathrm{Re}\,\alpha +i_{%
\mathbb{C}
}\mathrm{Im}\,\alpha \\ 
\mathrm{Re}\,\beta +i_{%
\mathbb{C}
}\mathrm{Im}\,\beta%
\end{array}%
\right) \text{.}
\end{equation}%
In \cite{lasenby93}, it was established a $1\leftrightarrow 1$ map between
Pauli column spinors and elements of the even subalgebra,%
\begin{equation}
\left\vert \psi \right\rangle =\left( 
\begin{array}{c}
a^{0}+i_{%
\mathbb{C}
}a^{3} \\ 
-a^{2}+i_{%
\mathbb{C}
}a^{1}%
\end{array}%
\right) \leftrightarrow \psi =a^{0}+a^{k}i\sigma _{k}\text{.}  \label{55}
\end{equation}%
where the coefficients $a^{0}$ and $a^{k}\in 
\mathbb{R}
$. Multivectors $\left\{ 1\text{, }i\sigma _{1}\text{, }i\sigma _{2}\text{, }%
i\sigma _{3}\right\} $ are the computational basis states of the real $4$%
-dimensional even subalgebra corresponding to the two-dimensional Hilbert
space $\mathcal{H}_{2}^{1}$ with standard computational basis given by $%
\mathcal{B}_{\mathcal{H}_{2}^{1}}\overset{\text{def}}{=}\left\{ \left\vert
0\right\rangle \text{, }\left\vert 1\right\rangle \right\} $. In the GA
formalism,%
\begin{equation}
\left\vert 0\right\rangle \leftrightarrow \psi _{\left\vert 0\right\rangle
}^{\left( \text{GA}\right) }\overset{\text{def}}{=}1\text{, }\left\vert
1\right\rangle \leftrightarrow \psi _{\left\vert 1\right\rangle }^{\left( 
\text{GA}\right) }\overset{\text{def}}{=}-i\sigma _{2}\text{.}
\end{equation}%
The action of the conventional quantum Pauli operators $\left\{ \hat{\Sigma}%
_{k}\text{, }i_{%
\mathbb{C}
}\hat{I}\right\} $ translates as \cite{lasenby93},%
\begin{equation}
\hat{\Sigma}_{k}\left\vert \psi \right\rangle \leftrightarrow \sigma
_{k}\psi \sigma _{3}\text{, }  \label{16}
\end{equation}%
with $k=1$, $2$, $3$ and,%
\begin{equation}
i_{%
\mathbb{C}
}\left\vert \psi \right\rangle \leftrightarrow \psi i\sigma _{3}\text{.}
\end{equation}%
In synthesis, in the single-particle theory, non-relativistic states are
constructed from the even subalgebra of the Pauli algebra with a basis
provided by the set $\left\{ 1\text{, }i\sigma _{k}\right\} $ with $k=1$, $2$%
, $3$. The role of the (single) imaginary unit of conventional quantum
theory is played by right multiplication by $i\sigma _{3}$. Verifying that
this translation scheme works properly is just a matter of simple
computations. Indeed, from (\ref{55}) and (\ref{16}) we obtain,%
\begin{eqnarray}
\hat{\Sigma}_{1}\left\vert \psi \right\rangle &=&\left( 
\begin{array}{c}
-a^{2}+i_{%
\mathbb{C}
}a^{1} \\ 
a^{0}+i_{%
\mathbb{C}
}a^{3}%
\end{array}%
\right) \leftrightarrow -a^{2}+a^{3}i\sigma _{1}-a^{0}i\sigma
_{2}+a^{1}i\sigma _{3}=\sigma _{1}\left( a^{0}+a^{k}i\sigma _{k}\right)
\sigma _{3}\text{,}  \notag \\
&&  \notag \\
\hat{\Sigma}_{2}\left\vert \psi \right\rangle &=&\left( 
\begin{array}{c}
a^{1}+i_{%
\mathbb{C}
}a^{2} \\ 
-a^{3}+i_{%
\mathbb{C}
}a^{0}%
\end{array}%
\right) \leftrightarrow a^{1}+a^{0}i\sigma _{1}+a^{3}i\sigma
_{2}+a^{2}i\sigma _{3}=\sigma _{2}\left( a^{0}+a^{k}i\sigma _{k}\right)
\sigma _{3}\text{,}  \notag \\
&&  \notag \\
\hat{\Sigma}_{3}\left\vert \psi \right\rangle &=&\left( 
\begin{array}{c}
a^{0}+i_{%
\mathbb{C}
}a^{3} \\ 
a^{2}-i_{%
\mathbb{C}
}a^{1}%
\end{array}%
\right) \leftrightarrow a^{0}-a^{1}i\sigma _{1}-a^{2}i\sigma
_{2}+a^{3}i\sigma _{3}=\sigma _{3}\left( a^{0}+a^{k}i\sigma _{k}\right)
\sigma _{3}\text{.}
\end{eqnarray}%
In the $n$-particle algebra there will be $n$-copies of $i\sigma _{3}$,
namely $i\sigma _{3}^{a}$ with $a=1$,.., $n$. However, in order to
faithfully mirror conventional quantum mechanics, the right-multiplication
by all of these must yield the same result. Therefore, it must be%
\begin{equation}
\psi i\sigma _{3}^{1}=\psi i\sigma _{3}^{2}=\text{...}=\psi i\sigma
_{3}^{n-1}=\psi i\sigma _{3}^{n}\text{.}  \label{ccn}
\end{equation}%
Relations in (\ref{ccn}) are obtained by introducing the $n$-particle
correlator $E_{n}$ defined as,%
\begin{equation}
E_{n}\overset{\text{def}}{=}\prod\limits_{b=2}^{n}\frac{1}{2}\left(
1-i\sigma _{3}^{1}i\sigma _{3}^{b}\right) \text{,}  \label{en}
\end{equation}%
and satisfying $E_{n}i\sigma _{3}^{a}=E_{n}i\sigma _{3}^{b}=J_{n}$; $\forall
a$, $b$. Notice that $E_{n}$ in (\ref{en}) has been defined by picking out
the $a=1$ space and correlating all the other spaces to this. However, the
value of $E_{n}$ is independent of which of the $n$ spaces is chosen and
correlated to. The complex structure is defined by $J_{n}=E_{n}i\sigma
_{3}^{a}$ with $J_{n}^{2}=-E_{n}$. Right-multiplication by the quantum
correlator $E_{n}$ is a projection operation that reduces the number of 
\emph{real} degrees of freedom from $4^{n}=\dim _{%
\mathbb{R}
}\left[ \mathfrak{cl}^{+}(3)\right] ^{n}$ to the expected $2^{n+1}=\dim _{%
\mathbb{R}
}\mathcal{H}_{2}^{n}$. The projection can be interpreted physically as
locking the phases of the various particles together. The \textit{reduced}
even subalgebra space will be denoted by $\left[ \mathfrak{cl}^{+}(3)\right]
^{n}/E_{n}$. Multivectors belonging to this space can be regarded as $n$%
-particle spinors (or, $n$-qubit states), analogous to $\mathfrak{cl}^{+}(3)$
for a single particle. In synthesis, the extension to multiparticle systems
involves a separate copy of the STA for each particle and the standard
imaginary unit induces correlations between these particle spaces.

\subsection{An Example: The $2$-Qubit Spacetime Algebra Formalism}

Quantum theory works with a single imaginary unit $i_{%
\mathbb{C}
}$, but in the $2$-particle algebra there are two bivectors playing the role
of $i_{%
\mathbb{C}
}$, $i\sigma _{3}^{1}$ and $i\sigma _{3}^{2}$. Right-multiplication of a
state by either of these has to result in the same state in order for the GA
treatment to faithfully mirror standard quantum mechanics. Therefore it must
be,%
\begin{equation}
\psi i\sigma _{3}^{1}=\psi i\sigma _{3}^{2}\text{.}  \label{a}
\end{equation}%
Manipulation of (\ref{a}) yields $\psi =\psi E$, where,%
\begin{equation}
E\overset{\text{def}}{=}\frac{1}{2}\left( 1-i\sigma _{3}^{1}i\sigma
_{3}^{2}\right) \text{, }E^{2}=E\text{.}  \label{EE}
\end{equation}%
Right-multiplication by $E$ is a projection operation. If we include this
factor on the right of all states, the number of \emph{real }degrees of
freedom decrease from $16$ to the expected $8$. The multivectorial basis $%
\mathcal{B}_{\mathfrak{cl}^{+}(3)\otimes \mathfrak{cl}^{+}(3)}$ spanning the 
$16$-dimensional geometric algebra $\mathfrak{cl}^{+}(3)\otimes \mathfrak{cl}%
^{+}(3)$ is given by,%
\begin{equation}
\mathcal{B}_{\mathfrak{cl}^{+}(3)\otimes \mathfrak{cl}^{+}(3)}\overset{\text{%
def}}{=}\left\{ 1\text{, }i\sigma _{l}^{1}\text{, }i\sigma _{k}^{2}\text{, }%
i\sigma _{l}^{1}i\sigma _{k}^{2}\text{ }\right\} \text{,}
\end{equation}%
with $k$ and $l=1$, $2$, $3$. Right-multiplying the multivectors in $%
\mathcal{B}_{\mathfrak{cl}^{+}(3)\otimes \mathfrak{cl}^{+}(3)}$ by the
quantum projection operator $E$, we obtain,%
\begin{equation}
\mathcal{B}_{\mathfrak{cl}^{+}(3)\otimes \mathfrak{cl}^{+}(3)}\overset{E}{%
\rightarrow }\mathcal{B}_{\mathfrak{cl}^{+}(3)\otimes \mathfrak{cl}^{+}(3)}E%
\overset{\text{def}}{=}\left\{ E\text{, }i\sigma _{l}^{1}E\text{, }i\sigma
_{k}^{2}E\text{, }i\sigma _{l}^{1}i\sigma _{k}^{2}E\text{ }\right\} \text{.}
\end{equation}%
After some straightforward algebra, it follows that,%
\begin{eqnarray}
E &=&-i\sigma _{3}^{1}i\sigma _{3}^{2}E\text{, }i\sigma _{1}^{2}E=-i\sigma
_{3}^{1}i\sigma _{2}^{2}E\text{, }i\sigma _{2}^{2}E=i\sigma _{3}^{1}i\sigma
_{1}^{2}E\text{, }i\sigma _{3}^{2}E=i\sigma _{3}^{1}E\text{,}  \notag \\
&&  \notag \\
\text{ }i\sigma _{1}^{1}E &=&-i\sigma _{2}^{1}i\sigma _{3}^{2}E\text{, }%
i\sigma _{1}^{1}i\sigma _{1}^{2}E=-i\sigma _{2}^{1}i\sigma _{2}^{2}E\text{, }%
i\sigma _{1}^{1}i\sigma _{2}^{2}E=i\sigma _{2}^{1}i\sigma _{1}^{2}E\text{, }%
i\sigma _{1}^{1}i\sigma _{3}^{2}E=i\sigma _{2}^{1}E\text{.}
\end{eqnarray}%
Therefore, a suitable basis for the $8$-dimensional \textit{reduced} even
subalgebra $\left[ \mathfrak{cl}^{+}(3)\otimes \mathfrak{cl}^{+}(3)\right]
/E $ is given by,%
\begin{equation}
\mathcal{B}_{\left[ \mathfrak{cl}^{+}(3)\otimes \mathfrak{cl}^{+}(3)\right]
/E}\overset{\text{def}}{=}\left\{ 1\text{, }i\sigma _{1}^{2}\text{, }i\sigma
_{2}^{2}\text{, }i\sigma _{3}^{2}\text{, }i\sigma _{1}^{1}\text{, }i\sigma
_{1}^{1}i\sigma _{1}^{2}\text{, }i\sigma _{1}^{1}i\sigma _{2}^{2}\text{, }%
i\sigma _{1}^{1}i\sigma _{3}^{2}\right\} \text{.}  \label{bb}
\end{equation}%
The basis in (\ref{bb}) spans $\left[ \mathfrak{cl}^{+}(3)\otimes \mathfrak{%
cl}^{+}(3)\right] /E$ and is the analog of a suitable standard complex basis
spanning the complex Hilbert space $\mathcal{H}_{2}^{2}$. The spacetime
algebra representation of a direct-product $2$-particle Pauli spinor (a $2$%
-qubits quantum state) is now given by $\psi ^{1}\phi ^{2}E$, where $\psi
^{1}$ and $\phi ^{2}$ are spinors (even multivectors) in their own spaces, $%
\left\vert \psi \text{, }\phi \right\rangle \leftrightarrow \psi ^{1}\phi
^{2}E$. A GA version of a standard complete basis for $2$-particle spin
states is provided by,%
\begin{eqnarray}
\left\vert 0\right\rangle \otimes \left\vert 0\right\rangle &=&\left( 
\begin{array}{c}
1 \\ 
0%
\end{array}%
\right) \otimes \left( 
\begin{array}{c}
1 \\ 
0%
\end{array}%
\right) \leftrightarrow E\text{, }\left\vert 0\right\rangle \otimes
\left\vert 1\right\rangle =\left( 
\begin{array}{c}
1 \\ 
0%
\end{array}%
\right) \otimes \left( 
\begin{array}{c}
0 \\ 
1%
\end{array}%
\right) \leftrightarrow -i\sigma _{2}^{2}E\text{,}  \notag \\
&&  \notag \\
\left\vert 1\right\rangle \otimes \left\vert 0\right\rangle &=&\left( 
\begin{array}{c}
0 \\ 
1%
\end{array}%
\right) \otimes \left( 
\begin{array}{c}
1 \\ 
0%
\end{array}%
\right) \leftrightarrow -i\sigma _{2}^{1}E\text{, }\left\vert 1\right\rangle
\otimes \left\vert 1\right\rangle =\left( 
\begin{array}{c}
0 \\ 
1%
\end{array}%
\right) \otimes \left( 
\begin{array}{c}
0 \\ 
1%
\end{array}%
\right) \leftrightarrow i\sigma _{2}^{1}i\sigma _{2}^{2}E\text{.}  \label{d2}
\end{eqnarray}%
For instance, a standard entangled state between a pair of $2$-level
systems, a spin singlet state is defined as,%
\begin{equation}
\left\vert \psi _{\text{singlet}}\right\rangle \overset{\text{def}}{=}\frac{1%
}{\sqrt{2}}\left\{ \left( 
\begin{array}{c}
1 \\ 
0%
\end{array}%
\right) \otimes \left( 
\begin{array}{c}
0 \\ 
1%
\end{array}%
\right) -\left( 
\begin{array}{c}
0 \\ 
1%
\end{array}%
\right) \otimes \left( 
\begin{array}{c}
1 \\ 
0%
\end{array}%
\right) \right\} =\frac{1}{\sqrt{2}}\left( \left\vert 01\right\rangle
-\left\vert 10\right\rangle \right) \text{.}  \label{d3}
\end{equation}%
From (\ref{EE}), (\ref{d2}) and (\ref{d3}), it follows that the GA\ analog
of $\left\vert \psi _{\text{singlet}}\right\rangle $ is given by,%
\begin{equation}
\mathcal{H}_{2}^{2}\ni \left\vert \psi _{\text{singlet}}\right\rangle
\leftrightarrow \psi _{\text{singlet}}^{\left( \text{GA}\right) }\in \left[ 
\mathfrak{cl}^{+}(3)\right] ^{2}\text{,}
\end{equation}%
where,%
\begin{equation}
\psi _{\text{singlet}}^{\left( \text{GA}\right) }=\frac{1}{2^{\frac{3}{2}}}%
\left( i\sigma _{2}^{1}-i\sigma _{2}^{2}\right) \left( 1-i\sigma
_{3}^{1}i\sigma _{3}^{2}\right) \text{.}
\end{equation}%
Furthermore, the role of multiplication by the quantum imaginary $i_{%
\mathbb{C}
}$ for $2$-particle states is taken by right-sided multiplication by $J$,%
\begin{equation}
J=Ei\sigma _{3}^{1}=Ei\sigma _{3}^{2}=\frac{1}{2}\left( i\sigma
_{3}^{1}+i\sigma _{3}^{2}\right) \text{,}
\end{equation}%
so that $J^{2}=-E$. The action of $2$-particle Pauli operators is the
following,%
\begin{equation}
\hat{\Sigma}_{k}\otimes \hat{I}\left\vert \psi \right\rangle \leftrightarrow
-i\sigma _{k}^{1}\psi J\text{, }\hat{\Sigma}_{k}\otimes \hat{\Sigma}%
_{l}\left\vert \psi \right\rangle \leftrightarrow -i\sigma _{k}^{1}i\sigma
_{l}^{2}\psi E\text{, }\hat{I}\otimes \hat{\Sigma}_{k}\left\vert \psi
\right\rangle \leftrightarrow -i\sigma _{k}^{2}\psi J\text{.}  \label{impo1}
\end{equation}%
For instance, the second Equation in (\ref{impo1}) follows from the
following line of reasoning,%
\begin{equation}
\hat{\Sigma}_{l}^{2}\left\vert \psi \right\rangle \leftrightarrow \sigma
_{l}^{2}\psi \sigma _{3}^{2}=\sigma _{l}^{2}\psi E\sigma _{3}^{2}=-\sigma
_{l}^{2}\psi Eii\sigma _{3}^{2}=-i\sigma _{l}^{2}\psi Ei\sigma
_{3}^{2}=-i\sigma _{l}^{2}\psi J\text{,}
\end{equation}%
and therefore,%
\begin{equation}
\hat{\Sigma}_{k}\otimes \hat{\Sigma}_{l}\left\vert \psi \right\rangle
\leftrightarrow \left( -i\sigma _{k}^{1}\right) \left( -i\sigma
_{l}^{2}\right) \psi J^{2}=-i\sigma _{k}^{1}i\sigma _{l}^{2}\psi E\text{.}
\end{equation}%
Finally, recalling that $i_{%
\mathbb{C}
}\hat{\Sigma}_{k}\left\vert \psi \right\rangle \leftrightarrow i\sigma
_{k}\psi $, we point out that,%
\begin{equation}
i_{%
\mathbb{C}
}\hat{\Sigma}_{k}\otimes \hat{I}\left\vert \psi \right\rangle
\leftrightarrow i\sigma _{k}^{1}\psi \text{ and, }\hat{I}\otimes i_{%
\mathbb{C}
}\hat{\Sigma}_{k}\left\vert \psi \right\rangle \leftrightarrow i\sigma
_{k}^{2}\psi \text{.}
\end{equation}%
More details on the MSTA\ formalism can be found in \cite{lasenby93, cd93,
doran96, somaroo99}.

\section{Geometric Algebra and Quantum Computation}

Interesting quantum computations may require constructions of complicated
computational networks with several gates acting on $n$-qubits defining a
non-trivial quantum algorithm. Therefore it is of great practical importance
finding a convenient \textit{universal} set of quantum gates. A set of
quantum gates $\left\{ \hat{U}_{i}\right\} $ is said to be \textit{universal}
if any logical operation $\hat{U}_{L}$ can be written as \cite{NIELSEN} ,%
\begin{equation}
\hat{U}_{L}=\prod\limits_{\hat{U}_{l}\subset \left\{ \hat{U}_{i}\right\} }%
\hat{U}_{l}\text{.}
\end{equation}%
In this Section, we present an explicit GA characterization of $1$ and $2$%
-qubit quantum states together with a GA characterization of a universal set
of quantum gates for quantum computation. Furthermore, we mention the
extension of the MSTA formalism to density matrices.

\subsection{Geometric Algebra and 1-Qubit Quantum Computing}

We consider simple circuit models of quantum computation with $1$-qubit
quantum gates in the GA formalism.

$\emph{Quantum}$ $\emph{NOT\ Gate}$ $\emph{(or}$ $\emph{Bit}$ $\emph{Flip}$ $%
\emph{Quantum}$ $\emph{Gate)}$. A nontrivial reversible operation we can
apply to a single qubit is the NOT operation (gate) denoted by the symbol $%
\hat{\Sigma}_{1}$. For the sake of simplicity, we will first study the
action of quantum gates in the GA formalism acting on $1$-qubit quantum
gates given by $\psi _{\left\vert q\right\rangle }^{\left( \text{GA}\right)
}=a^{0}+a^{2}i\sigma _{2}$. Then, $\hat{\Sigma}_{1}^{\left( \text{GA}\right)
}$ is defined as,%
\begin{equation}
\hat{\Sigma}_{1}\left\vert q\right\rangle \overset{\text{def}}{=}\left\vert
q\oplus 1\right\rangle \leftrightarrow \psi _{\left\vert q\oplus
1\right\rangle }^{\left( \text{GA}\right) }\overset{\text{def}}{=}\sigma
_{1}\left( a^{0}+a^{2}i\sigma _{2}\right) \sigma _{3}\text{.}
\end{equation}%
Recalling that the unit pseudoscalar $i\overset{\text{def}}{=}\sigma
_{1}\sigma _{2}\sigma _{3}$ is such that $i\sigma _{k}=\sigma _{k}i$ with $%
k=1$, $2$, $3$ and recalling the geometric product rule,%
\begin{equation}
\sigma _{i}\sigma _{j}=\sigma _{i}\cdot \sigma _{j}+\sigma _{i}\wedge \sigma
_{j}=\delta _{ij}+i\varepsilon _{ijk}\sigma _{k}\text{,}  \label{aa}
\end{equation}%
we obtain,%
\begin{equation}
\hat{\Sigma}_{1}\left\vert q\right\rangle \overset{\text{def}}{=}\left\vert
q\oplus 1\right\rangle \leftrightarrow \psi _{\left\vert q\oplus
1\right\rangle }^{\left( \text{GA}\right) }=-\left( a^{2}+a^{0}i\sigma
_{2}\right) \text{.}  \label{uno}
\end{equation}%
For the sake of completeness, we point out that the unitary quantum gate $%
\hat{\Sigma}_{1}^{\left( \text{GA}\right) }$ acts on the GA computational
basis states $\left\{ 1\text{, }i\sigma _{1}\text{, }i\sigma _{2}\text{, }%
i\sigma _{3}\right\} $ as follows,%
\begin{equation}
\hat{\Sigma}_{1}^{\left( \text{GA}\right) }:1\rightarrow -i\sigma _{2}\text{%
, }\hat{\Sigma}_{1}^{\left( \text{GA}\right) }:i\sigma _{1}\rightarrow
i\sigma _{3}\text{, }\hat{\Sigma}_{1}^{\left( \text{GA}\right) }:i\sigma
_{2}\rightarrow -1\text{, }\hat{\Sigma}_{1}^{\left( \text{GA}\right)
}:i\sigma _{3}\rightarrow i\sigma _{1}\text{.}
\end{equation}

\emph{Phase Flip Quantum Gate. }Another nontrivial reversible operation we
can apply to a single qubit is the phase flip gate denoted by the symbol $%
\hat{\Sigma}_{3}$. In the GA formalism, the action of the unitary quantum
gate $\hat{\Sigma}_{3}^{\left( \text{GA}\right) }$ on the multivector $\psi
_{\left\vert q\right\rangle }^{\left( \text{GA}\right) }=a^{0}+a^{2}i\sigma
_{2}$ is given by,%
\begin{equation}
\hat{\Sigma}_{3}\left\vert q\right\rangle \overset{\text{def}}{=}\left(
-1\right) ^{q}\left\vert q\right\rangle \leftrightarrow \psi _{\left(
-1\right) ^{q}\left\vert q\right\rangle }^{\left( \text{GA}\right) }\overset{%
\text{def}}{=}\sigma _{3}\left( a^{0}+a^{2}i\sigma _{2}\right) \sigma _{3}%
\text{.}
\end{equation}%
From (\ref{a}) and (\ref{aa}) it turns out that,%
\begin{equation}
\hat{\Sigma}_{3}\left\vert q\right\rangle \overset{\text{def}}{=}\left(
-1\right) ^{q}\left\vert q\right\rangle \leftrightarrow \psi _{\left(
-1\right) ^{q}\left\vert q\right\rangle }^{\left( \text{GA}\right)
}=a^{0}-a^{2}i\sigma _{2}\text{.}  \label{due}
\end{equation}%
Finally, the unitary quantum gate $\hat{\Sigma}_{3}^{\left( \text{GA}\right)
}$ acts on the GA computational basis states $\left\{ 1\text{, }i\sigma _{1}%
\text{, }i\sigma _{2}\text{, }i\sigma _{3}\right\} $ in the following manner,%
\begin{equation}
\hat{\Sigma}_{3}^{\left( \text{GA}\right) }:1\rightarrow 1\text{, }\hat{%
\Sigma}_{3}^{\left( \text{GA}\right) }:i\sigma _{1}\rightarrow -i\sigma _{1}%
\text{, }\hat{\Sigma}_{3}^{\left( \text{GA}\right) }:i\sigma _{2}\rightarrow
-i\sigma _{2}\text{, }\hat{\Sigma}_{3}^{\left( \text{GA}\right) }:i\sigma
_{3}\rightarrow i\sigma _{3}\text{.}
\end{equation}

\emph{Combined Bit and Phase Flip Quantum Gates. }A suitable combination of
the two reversible operations $\hat{\Sigma}_{1}$ and $\hat{\Sigma}_{3}$
gives rise to another nontrivial reversible operation we can apply to a
single qubit. Such operation is denoted by the symbol $\hat{\Sigma}_{2}$ $%
\overset{\text{def}}{=}i_{%
\mathbb{C}
}\hat{\Sigma}_{1}\circ \hat{\Sigma}_{3}$. The action of $\hat{\Sigma}%
_{2}^{\left( \text{GA}\right) }$ on $\psi _{\left\vert q\right\rangle
}^{\left( \text{GA}\right) }=a^{0}+a^{2}i\sigma _{2}$ is given by,%
\begin{equation}
\hat{\Sigma}_{2}\left\vert q\right\rangle \overset{\text{def}}{=}i_{%
\mathbb{C}
}\left( -1\right) ^{q}\left\vert q\oplus 1\right\rangle \leftrightarrow \psi
_{i_{%
\mathbb{C}
}\left( -1\right) ^{q}\left\vert q\oplus 1\right\rangle }^{\left( \text{GA}%
\right) }\overset{\text{def}}{=}\sigma _{2}\left( a^{0}+a^{2}i\sigma
_{2}\right) \sigma _{3}\text{.}
\end{equation}%
From (\ref{a}) and (\ref{aa}) it turns out that,%
\begin{equation}
\hat{\Sigma}_{2}\left\vert q\right\rangle \overset{\text{def}}{=}i_{%
\mathbb{C}
}\left( -1\right) ^{q}\left\vert q\oplus 1\right\rangle \leftrightarrow \psi
_{i_{%
\mathbb{C}
}\left( -1\right) ^{q}\left\vert q\oplus 1\right\rangle }^{\left( \text{GA}%
\right) }=\left( a^{2}-a^{0}i\sigma _{2}\right) i\sigma _{3}\text{.}
\end{equation}%
Indeed, using (\ref{aa}) and the fact that $i\sigma _{k}=\sigma _{k}i$ for $%
k=1$, $2$, $3$, we obtain, 
\begin{equation}
\sigma _{2}\left( a^{0}+a^{2}i\sigma _{2}\right) \sigma _{3}=\left(
a^{2}-a^{0}i\sigma _{2}\right) i\sigma _{3}\text{.}
\end{equation}%
Finally, the action of the unitary quantum gate $\hat{\Sigma}_{2}^{\left( 
\text{GA}\right) }$ on the GA computational basis states $\left\{ 1\text{, }%
i\sigma _{1}\text{, }i\sigma _{2}\text{, }i\sigma _{3}\right\} $ is,%
\begin{equation}
\hat{\Sigma}_{2}^{\left( \text{GA}\right) }:1\rightarrow i\sigma _{1}\text{, 
}\hat{\Sigma}_{2}^{\left( \text{GA}\right) }:i\sigma _{1}\rightarrow 1\text{%
, }\hat{\Sigma}_{2}^{\left( \text{GA}\right) }:i\sigma _{2}\rightarrow
i\sigma _{3}\text{, }\hat{\Sigma}_{2}^{\left( \text{GA}\right) }:i\sigma
_{3}\rightarrow i\sigma _{2}\text{.}
\end{equation}

\emph{Hadamard Quantum Gate. }The GA analog of the Walsh-Hadamard quantum
gate $\hat{H}$ $\overset{\text{def}}{=}\frac{\hat{\Sigma}_{1}+\hat{\Sigma}%
_{3}}{\sqrt{2}}$, $\hat{H}^{\left( \text{GA}\right) }$ acts on $\psi
_{\left\vert q\right\rangle }^{\left( \text{GA}\right) }=a^{0}+a^{2}i\sigma
_{2}$ as follows, 
\begin{equation}
\hat{H}\left\vert q\right\rangle \overset{\text{def}}{=}\frac{1}{\sqrt{2}}%
\left[ \left\vert q\oplus 1\right\rangle +\left( -1\right) ^{q}\left\vert
q\right\rangle \right] \leftrightarrow \psi _{\hat{H}\left\vert
q\right\rangle }^{\left( \text{GA}\right) }\overset{\text{def}}{=}\left( 
\frac{\sigma _{1}+\sigma _{3}}{\sqrt{2}}\right) \left( a^{0}+a^{2}i\sigma
_{2}\right) \sigma _{3}\text{.}  \label{tre}
\end{equation}%
Using (\ref{uno}) and (\ref{due}), (\ref{tre}) becomes,%
\begin{equation}
\hat{H}\left\vert q\right\rangle \overset{\text{def}}{=}\frac{1}{\sqrt{2}}%
\left[ \left\vert q\oplus 1\right\rangle +\left( -1\right) ^{q}\left\vert
q\right\rangle \right] \leftrightarrow \psi _{\hat{H}\left\vert
q\right\rangle }^{\left( \text{GA}\right) }=\frac{a^{0}}{\sqrt{2}}\left(
1-i\sigma _{2}\right) -\frac{a^{2}}{\sqrt{2}}\left( 1+i\sigma _{2}\right) 
\text{.}
\end{equation}%
Notice that GA versions of the Hadamard transformed computational states, $%
\left\vert +\right\rangle $ and $\left\vert -\right\rangle $, are given by,%
\begin{equation}
\left\vert +\right\rangle \overset{\text{def}}{=}\frac{\left\vert
0\right\rangle +\left\vert 1\right\rangle }{\sqrt{2}}\leftrightarrow \psi
_{\left\vert +\right\rangle }^{\left( \text{GA}\right) }=\frac{1-i\sigma _{2}%
}{\sqrt{2}}\text{ and, }\left\vert -\right\rangle \overset{\text{def}}{=}%
\frac{\left\vert 0\right\rangle -\left\vert 1\right\rangle }{\sqrt{2}}%
\leftrightarrow \psi _{\left\vert -\right\rangle }^{\left( \text{GA}\right)
}=\frac{1+i\sigma _{2}}{\sqrt{2}}\text{,}
\end{equation}%
respectively. Finally, the unitary quantum gate $\hat{H}^{\left( \text{GA}%
\right) }$ acts on the GA computational basis states $\left\{ 1\text{, }%
i\sigma _{1}\text{, }i\sigma _{2}\text{, }i\sigma _{3}\right\} $ as follows,%
\begin{equation}
\hat{H}^{\left( \text{GA}\right) }:1\rightarrow \frac{1-i\sigma _{2}}{\sqrt{2%
}}\text{, }\hat{H}^{\left( \text{GA}\right) }:i\sigma _{1}\rightarrow \frac{%
-i\sigma _{1}+i\sigma _{3}}{\sqrt{2}}\text{, }\hat{H}^{\left( \text{GA}%
\right) }:i\sigma _{2}\rightarrow -\frac{1+i\sigma _{2}}{\sqrt{2}}\text{, }%
\hat{H}^{\left( \text{GA}\right) }:i\sigma _{3}\rightarrow \frac{i\sigma
_{1}+i\sigma _{3}}{\sqrt{2}}\text{.}
\end{equation}

\emph{Rotation Gate. }The action of rotation gates $\hat{R}_{\theta
}^{\left( \text{GA}\right) }$ on $\psi _{\left\vert q\right\rangle }^{\left( 
\text{GA}\right) }=a^{0}+a^{2}i\sigma _{2}$ is defined as,%
\begin{equation}
\hat{R}_{\theta }\left\vert q\right\rangle \overset{\text{def}}{=}\left[ 
\frac{1+\exp \left( i_{%
\mathbb{C}
}\theta \right) }{2}+\left( -1\right) ^{q}\frac{1-\exp \left( i_{%
\mathbb{C}
}\theta \right) }{2}\right] \left\vert q\right\rangle \leftrightarrow \psi _{%
\hat{R}_{\theta }\left\vert q\right\rangle }^{\left( \text{GA}\right) }%
\overset{\text{def}}{=}a^{0}+a^{2}i\sigma _{2}\left( \cos \theta +i\sigma
_{3}\sin \theta \right) \text{.}
\end{equation}%
The unitary quantum gate $\hat{R}_{\theta }^{\left( \text{GA}\right) }$ acts
on the GA\ computational basis states $\left\{ 1\text{, }i\sigma _{1}\text{, 
}i\sigma _{2}\text{, }i\sigma _{3}\right\} $ as follows,%
\begin{equation}
\hat{R}_{\theta }^{\left( \text{GA}\right) }:1\rightarrow 1\text{, }\hat{R}%
_{\theta }^{\left( \text{GA}\right) }:i\sigma _{1}\rightarrow i\sigma
_{1}\left( \cos \theta +i\sigma _{3}\sin \theta \right) \text{, }\hat{R}%
_{\theta }^{\left( \text{GA}\right) }:i\sigma _{2}\rightarrow i\sigma
_{2}\left( \cos \theta +i\sigma _{3}\sin \theta \right) \text{, }\hat{R}%
_{\theta }^{\left( \text{GA}\right) }:i\sigma _{3}\rightarrow i\sigma _{3}%
\text{.}
\end{equation}

\emph{Phase Quantum Gate and }$\frac{\pi }{8}$\emph{-Quantum Gate. }The
phase gate $\hat{S}^{\left( \text{GA}\right) }$ acts on on $\psi
_{\left\vert q\right\rangle }^{\left( \text{GA}\right) }=a^{0}+a^{2}i\sigma
_{2}$ as follows,%
\begin{equation}
\hat{S}\left\vert q\right\rangle \overset{\text{def}}{=}\left[ \frac{1+i_{%
\mathbb{C}
}}{2}+\left( -1\right) ^{q}\frac{1-i_{%
\mathbb{C}
}}{2}\right] \left\vert q\right\rangle \leftrightarrow \psi _{\hat{S}%
\left\vert q\right\rangle }^{\left( \text{GA}\right) }\overset{\text{def}}{=}%
a^{0}+\left( a^{2}i\sigma _{2}\right) i\sigma _{3}\text{.}
\end{equation}%
Furthermore, the unitary quantum gate $\hat{S}^{\left( \text{GA}\right) }$
acts on the GA\ computational basis states $\left\{ 1\text{, }i\sigma _{1}%
\text{, }i\sigma _{2}\text{, }i\sigma _{3}\right\} $ as follows,%
\begin{equation}
\hat{S}^{\left( \text{GA}\right) }:1\rightarrow 1\text{, }\hat{S}^{\left( 
\text{GA}\right) }:i\sigma _{1}\rightarrow i\sigma _{2}\text{, }\hat{S}%
^{\left( \text{GA}\right) }:i\sigma _{2}\rightarrow -i\sigma _{1}\text{, }%
\hat{S}^{\left( \text{GA}\right) }:i\sigma _{3}\rightarrow i\sigma _{3}\text{%
.}
\end{equation}%
The GA analog of the $\frac{\pi }{8}$-quantum gate $\hat{T}$ is defined as,%
\begin{equation}
\hat{T}\left\vert q\right\rangle \overset{\text{def}}{=}\left[ \frac{1+\exp
\left( i_{%
\mathbb{C}
}\frac{\pi }{4}\right) }{2}+\left( -1\right) ^{q}\frac{1-\exp \left( i_{%
\mathbb{C}
}\frac{\pi }{4}\right) }{2}\right] \left\vert q\right\rangle \leftrightarrow
\psi _{\hat{T}\left\vert q\right\rangle }^{\left( \text{GA}\right) }\overset{%
\text{def}}{=}\frac{1}{\sqrt{2}}\left( a^{0}+a^{2}i\sigma _{2}\right) \left(
1+i\sigma _{3}\right) \text{.}
\end{equation}%
Finally, the unitary quantum gate $\hat{T}^{\left( \text{GA}\right) }$ acts
on the GA\ computational basis states $\left\{ 1\text{, }i\sigma _{1}\text{, 
}i\sigma _{2}\text{, }i\sigma _{3}\right\} $ as follows,%
\begin{equation}
\hat{T}^{\left( \text{GA}\right) }:1\rightarrow 1\text{, }\hat{T}^{\left( 
\text{GA}\right) }:i\sigma _{1}\rightarrow i\sigma _{1}\frac{\left(
1+i\sigma _{3}\right) }{\sqrt{2}}\text{, }\hat{T}^{\left( \text{GA}\right)
}:i\sigma _{2}\rightarrow i\sigma _{2}\frac{\left( 1+i\sigma _{3}\right) }{%
\sqrt{2}}\text{, }\hat{T}^{\left( \text{GA}\right) }:i\sigma _{3}\rightarrow
i\sigma _{3}\text{.}
\end{equation}%
In conclusion, the action of some of the most relevant $1$-qubit quantum
gates in the GA formalism on the GA computational basis states $\left\{ 1%
\text{, }i\sigma _{1}\text{, }i\sigma _{2}\text{, }i\sigma _{3}\right\} $
can be summarized in the following tabular form:%
\begin{equation}
\begin{tabular}{||c||c||c||c||c||c||c||}
\hline\hline
$1$-\emph{Qubit States} & \emph{NOT} & \emph{Phase Flip} & \emph{Bit and
Phase Flip} & \emph{Hadamard} & \emph{Rotation} & $\frac{\pi }{8}$\emph{-Gate%
} \\ \hline\hline
$1$ & $-i\sigma _{2}$ & $1$ & $i\sigma _{1}$ & $\frac{1-i\sigma _{2}}{\sqrt{2%
}}$ & $1$ & $1$ \\ \hline\hline
$i\sigma _{1}$ & $i\sigma _{3}$ & $-i\sigma _{1}$ & $1$ & $\frac{-i\sigma
_{1}+i\sigma _{3}}{\sqrt{2}}$ & $i\sigma _{1}\left( \cos \theta +i\sigma
_{3}\sin \theta \right) $ & $i\sigma _{1}\frac{\left( 1+i\sigma _{3}\right) 
}{\sqrt{2}}$ \\ \hline\hline
$i\sigma _{2}$ & $-1$ & $-i\sigma _{2}$ & $i\sigma _{3}$ & $-\frac{1+i\sigma
_{2}}{\sqrt{2}}$ & $i\sigma _{2}\left( \cos \theta +i\sigma _{3}\sin \theta
\right) $ & $i\sigma _{2}\frac{\left( 1+i\sigma _{3}\right) }{\sqrt{2}}$ \\ 
\hline\hline
$i\sigma _{3}$ & $i\sigma _{1}$ & $i\sigma _{3}$ & $i\sigma _{2}$ & $\frac{%
i\sigma _{1}+i\sigma _{3}}{\sqrt{2}}$ & $i\sigma _{3}$ & $i\sigma
_{3}\rightarrow i\sigma _{3}$ \\ \hline\hline
\end{tabular}%
\text{.}  \label{T1}
\end{equation}%
Therefore, in the GA approach qubits become elements of the even subalgebra,
unitary quantum gates become rotors and the conventional complex structure
of quantum mechanics is controlled by the bivector $i\sigma _{3}$.

Quantum gates have a geometrical interpretation when expressed in the GA
formalism. Recall that in the conventional approach to quantum gates, an
arbitrary unitary operator on a single qubit can be written as a combination
of rotations together with global phase shifts on the qubit, $\hat{U}=e^{i_{%
\mathbb{C}
}\alpha }R_{\hat{n}}\left( \theta \right) $ for some\textbf{\ }\emph{real}%
\textbf{\ }numbers $\alpha $ and $\theta $ and a\textbf{\ }\emph{real}
three-dimensional unit vector $\hat{n}\equiv \left( n_{1}\text{, }n_{2}\text{%
, }n_{3}\right) $\textbf{.} For instance, the Hadamard gate $\hat{H}$ acting
on a single qubit has the properties\textbf{s} $\hat{H}\hat{\Sigma}_{1}\hat{H%
}=\hat{\Sigma}_{3}$ and $\hat{H}\hat{\Sigma}_{3}\hat{H}=\hat{\Sigma}_{1}$
with $\hat{H}^{2}=\hat{I}$. Therefore, $\hat{H}$ can be envisioned (up to an
overall phase) as a $\theta =\pi $ rotation about the axis $\hat{n}=\frac{1}{%
\sqrt{2}}\left( \hat{n}_{1}+\hat{n}_{3}\right) $ that rotates $\hat{x}$ to $%
\hat{z}$ and viceversa, $\hat{H}=-i_{%
\mathbb{C}
}R_{\frac{1}{\sqrt{2}}\left( \hat{n}_{1}+\hat{n}_{3}\right) }\left( \pi
\right) $\textbf{. }In GA, rotations are handled by means of rotors. The
Hadamard gate, for instance, has a simple\textbf{\ }\emph{real}\textbf{\emph{%
\ }}(no use of \emph{complex} numbers is needed) geometric interpretation:
it is represented by a rotor $\hat{H}^{\left( \text{GA}\right) }=e^{-i\frac{%
\pi }{2}\frac{\sigma _{1}+\sigma _{3}}{\sqrt{2}}}$ \ describing a rotation
by $\pi $ about the $\frac{\sigma _{1}+\sigma _{3}}{\sqrt{2}}$ axis.\textbf{%
\ }It is straightforward to show that the action of the rotor $\hat{H}%
^{\left( \text{GA}\right) }$ on the $1$\textbf{-}qubit computational basis
states satisfies (up to an overall irrelevant phase shift) the
transformation laws appearing in (\ref{T1}). We point out that when the
Hadamard gate is represented by a rotor for a rotation by $\pi $, $\hat{H}%
^{\left( \text{GA}\right) 2}=-1$. \ Therefore, it seems that the gate is
more accurately represented by a reflection rather than a rotation. The
phase difference may be important when state amplitudes transformed by the
Hadamard gate are combined with the ones transformed by other gates. In \cite%
{doran1}, it was also proposed treating the Hadamard gate as a rotation but
it is now recognized the problem with this interpretation. Similar geometric
considerations could be carried out for the other $1$-qubit gates \cite%
{doran1}.

\subsection{Geometric Algebra and $2$-Qubit Quantum Computing}

We consider simple circuit models of quantum computation with $2$-qubit
quantum gates in the GA formalism. Before doing so, we present an explicit
MSTA description of quantum Bell states.

\emph{Geometric Algebra and Bell States. }We present a GA characterization
of the set of maximally entangled $2$-qubits Bell states. Bell states are an
important example of maximally entangled quantum states and form an
orthonormal basis $\mathcal{B}_{\text{Bell}}$ in the product Hilbert space $%
\mathbb{C}
^{2}\otimes 
\mathbb{C}
^{2}\cong 
\mathbb{C}
^{4}$. Consider the $2$-qubit computational basis $\mathcal{B}_{\text{%
computational}}=\left\{ \left\vert 00\right\rangle \text{, }\left\vert
01\right\rangle \text{, }\left\vert 10\right\rangle \text{, }\left\vert
11\right\rangle \right\} $, then the four Bell states can be constructed as
follows \cite{NIELSEN},%
\begin{align}
\left\vert 0\right\rangle \otimes \left\vert 0\right\rangle & \rightarrow
\left\vert \psi _{\text{Bell}_{1}}\right\rangle \overset{\text{def}}{=}\left[
\hat{U}_{\text{CNOT}}\circ \left( \hat{H}\otimes \hat{I}\right) \right]
\left( \left\vert 0\right\rangle \otimes \left\vert 0\right\rangle \right) =%
\frac{1}{\sqrt{2}}\left( \left\vert 0\right\rangle \otimes \left\vert
0\right\rangle +\left\vert 1\right\rangle \otimes \left\vert 1\right\rangle
\right) \text{,}  \notag \\
&  \notag \\
\left\vert 0\right\rangle \otimes \left\vert 1\right\rangle & \rightarrow
\left\vert \psi _{\text{Bell}_{2}}\right\rangle \overset{\text{def}}{=}\left[
\hat{U}_{\text{CNOT}}\circ \left( \hat{H}\otimes \hat{I}\right) \right]
\left( \left\vert 0\right\rangle \otimes \left\vert 1\right\rangle \right) =%
\frac{1}{\sqrt{2}}\left( \left\vert 0\right\rangle \otimes \left\vert
1\right\rangle +\left\vert 1\right\rangle \otimes \left\vert 0\right\rangle
\right) \text{,}  \notag \\
&  \notag \\
\left\vert 1\right\rangle \otimes \left\vert 0\right\rangle & \rightarrow
\left\vert \psi _{\text{Bell}_{3}}\right\rangle \overset{\text{def}}{=}\left[
\hat{U}_{\text{CNOT}}\circ \left( \hat{H}\otimes \hat{I}\right) \right]
\left( \left\vert 1\right\rangle \otimes \left\vert 0\right\rangle \right) =%
\frac{1}{\sqrt{2}}\left( \left\vert 0\right\rangle \otimes \left\vert
0\right\rangle -\left\vert 1\right\rangle \otimes \left\vert 1\right\rangle
\right) \text{,}  \notag \\
&  \notag \\
\left\vert 1\right\rangle \otimes \left\vert 1\right\rangle & \rightarrow
\left\vert \psi _{\text{Bell}_{4}}\right\rangle \overset{\text{def}}{=}\left[
\hat{U}_{\text{CNOT}}\circ \left( \hat{H}\otimes \hat{I}\right) \right]
\left( \left\vert 1\right\rangle \otimes \left\vert 1\right\rangle \right) =%
\frac{1}{\sqrt{2}}\left( \left\vert 0\right\rangle \otimes \left\vert
1\right\rangle -\left\vert 1\right\rangle \otimes \left\vert 0\right\rangle
\right) \text{.}  \label{so}
\end{align}%
In (\ref{so}), $\hat{H}$ is the Hadamard gate and $\hat{U}_{\text{CNOT}}$ is
the CNOT gate. The Bell basis in $%
\mathbb{C}
^{2}\otimes 
\mathbb{C}
^{2}\cong 
\mathbb{C}
^{4}$ is given by,%
\begin{equation}
\mathcal{B}_{\text{Bell}}\overset{\text{def}}{=}\left\{ \left\vert \psi _{%
\text{Bell}_{1}}\right\rangle \text{, }\left\vert \psi _{\text{Bell}%
_{2}}\right\rangle \text{, }\left\vert \psi _{\text{Bell}_{3}}\right\rangle 
\text{, }\left\vert \psi _{\text{Bell}_{4}}\right\rangle \right\} \text{,}
\end{equation}%
where from (\ref{so}) we get,%
\begin{equation}
\left\vert \psi _{\text{Bell}_{1}}\right\rangle =\frac{1}{\sqrt{2}}\left( 
\begin{array}{c}
1 \\ 
0 \\ 
0 \\ 
1%
\end{array}%
\right) \text{, }\left\vert \psi _{\text{Bell}_{2}}\right\rangle =\frac{1}{%
\sqrt{2}}\left( 
\begin{array}{c}
0 \\ 
1 \\ 
1 \\ 
0%
\end{array}%
\right) \text{, }\left\vert \psi _{\text{Bell}_{3}}\right\rangle =\frac{1}{%
\sqrt{2}}\left( 
\begin{array}{c}
1 \\ 
0 \\ 
0 \\ 
-1%
\end{array}%
\right) \text{, }\left\vert \psi _{\text{Bell}_{4}}\right\rangle =\frac{1}{%
\sqrt{2}}\left( 
\begin{array}{c}
0 \\ 
1 \\ 
-1 \\ 
0%
\end{array}%
\right) \text{.}
\end{equation}%
From (\ref{d2}) and (\ref{so}), it follows that the GA version of Bell
states is given by,%
\begin{eqnarray}
\left\vert \psi _{\text{Bell}_{1}}\right\rangle &\leftrightarrow &\psi _{%
\text{Bell}_{1}}^{\left( \text{GA}\right) }=\frac{1}{2^{\frac{3}{2}}}\left(
1+i\sigma _{2}^{1}i\sigma _{2}^{2}\right) \left( 1-i\sigma _{3}^{1}i\sigma
_{3}^{2}\right) \text{, }\left\vert \psi _{\text{Bell}_{2}}\right\rangle
\leftrightarrow \psi _{\text{Bell}_{2}}^{\left( \text{GA}\right) }=-\frac{1}{%
2^{\frac{3}{2}}}\left( i\sigma _{2}^{1}+i\sigma _{2}^{2}\right) \left(
1-i\sigma _{3}^{1}i\sigma _{3}^{2}\right) \text{,}  \notag \\
&&  \notag \\
\left\vert \psi _{\text{Bell}_{3}}\right\rangle &\leftrightarrow &\psi _{%
\text{Bell}_{3}}^{\left( \text{GA}\right) }=\frac{1}{2^{\frac{3}{2}}}\left(
1-i\sigma _{2}^{1}i\sigma _{2}^{2}\right) \left( 1-i\sigma _{3}^{1}i\sigma
_{3}^{2}\right) \text{, }\left\vert \psi _{\text{Bell}_{4}}\right\rangle
\leftrightarrow \psi _{\text{Bell}_{4}}^{\left( \text{GA}\right) }=\frac{1}{%
2^{\frac{3}{2}}}\left( i\sigma _{2}^{1}-i\sigma _{2}^{2}\right) \left(
1-i\sigma _{3}^{1}i\sigma _{3}^{2}\right) \text{.}
\end{eqnarray}%
We point out that within the MSTA formalism, there is no need either for an
abstract spin space (the complex Hilbert space $\mathcal{H}_{2}^{n}$),
containing objects which have to be operated on by quantum unitary operators
(for instance, in the Bell states example, such operators are the CNOT
gates), or for an abstract index convention. The requirement for an explicit
matrix representation is also avoided. Within the MSTA formalism, the role
of operators is taken over by right or left multiplication by elements from
the same geometric algebra as spinors (qubits) are taken from. This is an
additional manifestation of a conceptual unification provided by GA - \
"spin (or qubit) space" and "unitary operators upon spin space" become
united, with both being just multivectors in real space. Indeed, such a
manifestation is not a special feature of our work since it appears in most
geometric algebra applications to classical and quantum mathematical physics.

\emph{CNOT Quantum Gate. }An convenient way to describe the CNOT quantum
gate is the following \cite{NIELSEN},%
\begin{equation}
\hat{U}_{\text{CNOT}}^{12}=\frac{1}{2}\left[ \left( \hat{I}^{1}+\hat{\Sigma}%
_{3}^{1}\right) \otimes \hat{I}^{2}+\left( \hat{I}^{1}-\hat{\Sigma}%
_{3}^{1}\right) \otimes \hat{\Sigma}_{1}^{2}\right] \text{,}
\end{equation}%
where $\hat{U}_{\text{CNOT}}^{12}$ is the CNOT gate from qubit $1$ to qubit $%
2$. It then follows that,%
\begin{equation}
\hat{U}_{\text{CNOT}}^{12}\left\vert \psi \right\rangle =\frac{1}{2}\left( 
\hat{I}^{1}\otimes \hat{I}^{2}+\hat{\Sigma}_{3}^{1}\otimes \hat{I}^{2}+\hat{I%
}^{1}\otimes \hat{\Sigma}_{1}^{2}-\hat{\Sigma}_{3}^{1}\otimes \hat{\Sigma}%
_{1}^{2}\right) \left\vert \psi \right\rangle \text{.}  \label{11}
\end{equation}%
From (\ref{impo1}) and (\ref{11}), we obtain%
\begin{equation}
\hat{I}^{1}\otimes \hat{I}^{2}\left\vert \psi \right\rangle \leftrightarrow
\psi \text{, }\hat{\Sigma}_{3}^{1}\otimes \hat{I}^{2}\left\vert \psi
\right\rangle \leftrightarrow -i\sigma _{3}^{1}\psi J\text{, }\hat{I}%
^{1}\otimes \hat{\Sigma}_{1}^{2}\left\vert \psi \right\rangle
\leftrightarrow -i\sigma _{1}^{2}\psi J\text{, }-\hat{\Sigma}_{3}^{1}\otimes 
\hat{\Sigma}_{1}^{2}\left\vert \psi \right\rangle \leftrightarrow i\sigma
_{3}^{1}i\sigma _{1}^{2}\psi E\text{.}  \label{12}
\end{equation}%
Finally, from (\ref{11}) and (\ref{12}), we get the GA version of the CNOT
gate,%
\begin{equation}
\hat{U}_{\text{CNOT}}^{12}\left\vert \psi \right\rangle \leftrightarrow 
\frac{1}{2}\left( \psi -i\sigma _{3}^{1}\psi J-i\sigma _{1}^{2}\psi
J+i\sigma _{3}^{1}i\sigma _{1}^{2}\psi E\right) \text{.}  \label{13}
\end{equation}

\emph{Controlled-Phase Gate.}The action of $\hat{U}_{\text{CP}}^{12}$ on $%
\left\vert \psi \right\rangle \in \mathcal{H}_{2}^{2}$ is given by \cite%
{NIELSEN},%
\begin{equation}
\hat{U}_{\text{CP}}^{12}\left\vert \psi \right\rangle =\frac{1}{2}\left[ 
\hat{I}^{1}\otimes \hat{I}^{2}+\hat{\Sigma}_{3}^{1}\otimes \hat{I}^{2}+\hat{I%
}^{1}\otimes \hat{\Sigma}_{3}^{2}-\hat{\Sigma}_{3}^{1}\otimes \hat{\Sigma}%
_{3}^{2}\right] \left\vert \psi \right\rangle \text{.}  \label{11a}
\end{equation}%
From (\ref{impo1}) and (\ref{11a}), we obtain%
\begin{equation}
\hat{I}^{1}\otimes \hat{I}^{2}\left\vert \psi \right\rangle \leftrightarrow
\psi \text{, }\hat{\Sigma}_{3}^{1}\otimes \hat{I}^{2}\left\vert \psi
\right\rangle \leftrightarrow -i\sigma _{3}^{1}\psi J\text{, }\hat{I}%
^{1}\otimes \hat{\Sigma}_{3}^{2}\left\vert \psi \right\rangle
\leftrightarrow -i\sigma _{3}^{2}\psi J\text{, }-\hat{\Sigma}_{3}^{1}\otimes 
\hat{\Sigma}_{3}^{2}\left\vert \psi \right\rangle \leftrightarrow i\sigma
_{3}^{1}i\sigma _{3}^{2}\psi E\text{.}  \label{11b}
\end{equation}%
Finally, from (\ref{11a}) and (\ref{11b}), we obtain the GA version of the
controlled-phase quantum gate,%
\begin{equation}
\hat{U}_{\text{CP}}^{12}\left\vert \psi \right\rangle \leftrightarrow \frac{1%
}{2}\left( \psi -i\sigma _{3}^{1}\psi J-i\sigma _{3}^{2}\psi J+i\sigma
_{3}^{1}i\sigma _{3}^{2}\psi E\right) \text{.}  \label{14}
\end{equation}

\emph{SWAP Gate. }The action of $\hat{U}_{\text{SWAP}}^{12}$ on $\left\vert
\psi \right\rangle \in \mathcal{H}_{2}^{2}$ is given by \cite{NIELSEN},%
\begin{equation}
\hat{U}_{\text{SWAP}}^{12}\left\vert \psi \right\rangle =\frac{1}{2}\left( 
\hat{I}^{1}\otimes \hat{I}^{2}+\hat{\Sigma}_{1}^{1}\otimes \hat{\Sigma}%
_{1}^{2}+\hat{\Sigma}_{2}^{1}\otimes \hat{\Sigma}_{2}^{2}+\hat{\Sigma}%
_{3}^{1}\otimes \hat{\Sigma}_{3}^{2}\right) \left\vert \psi \right\rangle 
\text{.}  \label{12a}
\end{equation}%
From (\ref{impo1}) and (\ref{12a}), we obtain%
\begin{equation}
\hat{I}^{1}\otimes \hat{I}^{2}\left\vert \psi \right\rangle \leftrightarrow
\psi \text{, }\hat{\Sigma}_{1}^{1}\otimes \hat{\Sigma}_{1}^{2}\left\vert
\psi \right\rangle \leftrightarrow -i\sigma _{1}^{1}i\sigma _{1}^{2}\psi E%
\text{, }\hat{\Sigma}_{2}^{1}\otimes \hat{\Sigma}_{2}^{2}\left\vert \psi
\right\rangle \leftrightarrow -i\sigma _{2}^{1}i\sigma _{2}^{2}\psi E\text{, 
}\hat{\Sigma}_{3}^{1}\otimes \hat{\Sigma}_{3}^{2}\left\vert \psi
\right\rangle \leftrightarrow -i\sigma _{3}^{1}i\sigma _{3}^{2}\psi E\text{.}
\label{12b}
\end{equation}%
Finally, from (\ref{12a}) and (\ref{12b}), we obtain the GA version of the
SWAP gate,%
\begin{equation}
\hat{U}_{\text{SWAP}}^{12}\left\vert \psi \right\rangle \leftrightarrow 
\frac{1}{2}\left( \psi -i\sigma _{1}^{1}i\sigma _{1}^{2}\psi E-i\sigma
_{2}^{1}i\sigma _{2}^{2}\psi E-i\sigma _{3}^{1}i\sigma _{3}^{2}\psi E\right) 
\text{.}  \label{15}
\end{equation}%
In conclusion, the action of some of the most relevant $2$-qubit quantum
gates in the GA formalism on the GA computational basis $\mathcal{B}_{\left[ 
\mathfrak{cl}^{+}(3)\otimes \mathfrak{cl}^{+}(3)\right] /E}$ can be
summarized in the following tabular form:%
\begin{equation}
\begin{tabular}{||c||c||c||}
\hline\hline
$2$-\emph{Qubit Gates} & $2$-\emph{Qubit States} & \emph{GA} \emph{Action of
Gates on States} \\ \hline\hline
CNOT & $\psi $ & $\frac{1}{2}\left( \psi -i\sigma _{3}^{1}\psi J-i\sigma
_{1}^{2}\psi J+i\sigma _{3}^{1}i\sigma _{1}^{2}\psi E\right) $ \\ 
\hline\hline
Controlled-Phase Gate & $\psi $ & $\frac{1}{2}\left( \psi -i\sigma
_{3}^{1}\psi J-i\sigma _{3}^{2}\psi J+i\sigma _{3}^{1}i\sigma _{3}^{2}\psi
E\right) $ \\ \hline\hline
SWAP & $\psi $ & $\frac{1}{2}\left( \psi -i\sigma _{1}^{1}i\sigma
_{1}^{2}\psi E-i\sigma _{2}^{1}i\sigma _{2}^{2}\psi E-i\sigma
_{3}^{1}i\sigma _{3}^{2}\psi E\right) $ \\ \hline\hline
\end{tabular}%
\text{.}
\end{equation}%
Two-qubit quantum gates have a geometric interpretation in terms of
rotations as well. For instance, the CNOT gate describes a rotation in one
qubit space\textbf{\ }\emph{conditional}\textbf{\ }on the state of another
qubit it is correlated with. The general expression of the corresponding
operator in GA is given by\textbf{\ }$\left( \hat{U}_{\text{CNOT}%
}^{12}\right) ^{\left( \text{GA}\right) }=e^{-i\frac{\pi }{2}\frac{1}{2}%
\sigma _{1}^{1}\left( 1-\sigma _{3}^{2}\right) }$\textbf{. }This operator
rotates the first qubit by $\pi $ about the axis $\sigma _{1}^{1}$ in those $%
2$-qubit states where the second qubit is along the $-\sigma _{2}^{3}$ axis.
Similar geometric considerations could be considered for the other $2$-qubit
gates \cite{somaroo}.

\subsection{Geometric Algebra and Density Operators}

For the sake of completeness, we point out that statistical aspects of
quantum systems cannot be described in terms of a single wavefunction.
Instead, they can be properly handled in terms of density matrices. The
density matrix for a pure state is given by,%
\begin{equation}
\hat{\rho}_{\text{pure}}=\left\vert \psi \right\rangle \left\langle \psi
\right\vert =\left( 
\begin{array}{cc}
\alpha \alpha ^{\ast } & \alpha \beta ^{\ast } \\ 
\beta \alpha ^{\ast } & \beta \beta ^{\ast }%
\end{array}%
\right) \text{.}
\end{equation}%
The expectation value of any observable $\hat{O}$ associated with the state $%
\left\vert \psi \right\rangle $ can be obtained from $\hat{\rho}_{\text{pure}%
}$ by writing $\left\langle \psi |\hat{O}|\psi \right\rangle =$Tr$\left( 
\hat{\rho}_{\text{pure}}\hat{O}\right) $. The GA version of $\hat{\rho}_{%
\text{pure}}$ is,%
\begin{equation}
\hat{\rho}_{\text{pure}}\rightarrow \rho _{\text{pure}}^{\left( \text{GA}%
\right) }=\psi \frac{1}{2}\left( 1+\sigma _{3}\right) \psi ^{\dagger }=\frac{%
1}{2}\left( 1+s\right) \text{,}
\end{equation}%
where $s\overset{\text{def}}{=}\psi \sigma _{3}\psi ^{\dagger }$ is the spin
vector \cite{doran96}. From a geometric point of view, $\rho _{\text{pure}%
}^{\left( \text{GA}\right) }$ is just the sum of a scalar and a vector. The
density matrix for a mixed state $\hat{\rho}_{\text{mixed}}$ is the weighted
sum of the density matrices for the pure states,%
\begin{equation}
\hat{\rho}_{\text{mixed}}=\sum_{j=1}^{n}\hat{\rho}_{j}=\sum_{j=1}^{n}p_{j}%
\left\vert \psi _{j}\right\rangle \left\langle \psi _{j}\right\vert \text{,}
\end{equation}%
with $p_{j}\in 
\mathbb{R}
$ for $j=1$,..., $n$ and $p_{1}+$...$+p_{n}=1$\textbf{. }In the GA\
formalism, addition is well-defined and the geometric algebra version of $%
\hat{\rho}_{\text{mixed}}$ becomes the sum,%
\begin{equation}
\hat{\rho}_{\text{mixed}}\rightarrow \rho _{\text{mixed}}^{\left( \text{GA}%
\right) }=\frac{1}{2}\sum_{j=1}^{n}\left( p_{j}+p_{j}s_{j}\right) =\frac{1}{2%
}\left( 1+P\right) \text{,}
\end{equation}%
where $P$ is the ensemble-average polarization vector (average spin vector)
with length $\left\Vert P\right\Vert \leq 1$. The magnitude of $P$ measures
the degree of alignment among the unit length polarization vectors of the
individual numbers of the ensemble. We point out that $\rho _{\text{mixed}%
}^{\left( \text{GA}\right) }$ is the geometric algebra expression of the
density operator of an ensemble of identical and non-interacting qubits.
More generally, we could also consider density operators of interacting
multi-qubit systems. The MSTA version of the density matrix of $n$%
-interacting qubits reads,%
\begin{equation}
\rho _{\text{multi-qubit}}^{\left( \text{GA}\right) }=\overline{\left( \psi
E_{n}\right) E_{+}\left( \psi E_{n}\right) ^{\sim }}\text{,}
\end{equation}%
where $E_{n}$ is the $n$-particle correlator and\textbf{\ }$E_{+}\overset{%
\text{def}}{=}E_{+}^{1}E_{+}^{2}$...$E_{+}^{n}$ is the geometric product of $%
n$-idempotents with $E_{\pm }^{k}=\frac{1\pm \sigma _{3}^{k}}{2}$ and $k=1$%
,..., $n$. The symbol tilde denotes the space-time reverse and the over-line
denotes the ensemble-average. A more detailed application of the GA
formalism to general density matrices appears in \cite{doran1}.

\section{On the universality of quantum gates and geometric algebra}

In this Section, using the above mentioned explicit characterization and the
GA description of the Lie algebras $SO\left( 3\right) $ and $SU\left(
2\right) $ based on the rotor group $\emph{Spin}^{+}\left( 3\text{, }%
0\right) $ formalism, we reexamine Boykin's proof of universality of quantum
gates. In the first part, we introduce the rotor group. In the second part,
we introduce few universal sets of quantum gates. In the last part, we
present our GA-based proof.

\subsection{On $SO(3)$, $SU(2)$ and the Rotor Group in Geometric Algebra}

Since Boykin's proof heavily relies on the properties of rotations in
three-dimensional space and on the local isomorphism between $SO(3)$ and $%
SU(2)$, we briefly present the the GA description of such Lie groups via the
rotor group $\emph{Spin}^{+}\left( 3\text{, }0\right) $.

\subsubsection{Remarks on $SO(3)$ and $SU(2)$}

Two important groups in physics are the $3$-dimensional Lie groups $SO(3)$
with Lie algebra $\mathfrak{so(}3\mathfrak{)}$ and $SU(2)$ with Lie algebra $%
\mathfrak{su(}2\mathfrak{)}$ \cite{morton}. The former is the group of
rotations of three-dimensional space, i.e. the group of orthogonal
transformations with determinant $1$,%
\begin{equation}
SO(3)\overset{\text{def}}{=}\left\{ M\in GL(3\text{, }%
\mathbb{R}
):MM^{t}=M^{t}M=I_{3\times 3}\text{, }\det M=1\right\} \text{,}
\end{equation}%
where "$t$" denotes the transpose of a matrix and $GL(3$, $%
\mathbb{R}
)$ is the set of non-singular linear transformations in $%
\mathbb{R}
^{3}$ which are represented by $3\times 3$ non singular matrices with real
entries. The latter is the group of all $2\times 2$ unitary complex matrices
with determinant equal to $1$,%
\begin{equation}
SU(2)\overset{\text{def}}{=}\left\{ M\in GL(2\text{, }%
\mathbb{C}
):MM^{\dagger }=M^{\dagger }M=I_{2\times 2}\text{, }\det M=1\right\} \text{,}
\end{equation}%
where "$\dagger $" denotes the Hermitian conjugate and $GL(2$, $%
\mathbb{C}
)$ is the set of non-singular linear transformations in $%
\mathbb{C}
^{2}$ which are represented by $2\times 2$ non singular matrices with
complex entries. The Lie algebras $\mathfrak{so(}3\mathfrak{)}$ and $%
\mathfrak{su(}2\mathfrak{)}$ are isomorphic, $\mathfrak{so(}3\mathfrak{)}%
\cong \mathfrak{su(}2\mathfrak{)}$. The Lie groups $SO(3)$ and $SU(2)$ are 
\emph{locally} isomorphic, they are indistinguishable at the level of
infinitesimal transformations. However, they differ at a global level, i. e.
far from identity. This means that $SO(3)$ and $SU(2)$ are not isomorphic.
In $SO(3)$ a rotation by $2\pi $ is the same as the identity. Instead, $%
SU(2) $ is periodic only under rotations by $4\pi $. This means that an
object that picks a minus sign under a rotation by $2\pi $ is an acceptable
representation of $SU(2)$, while it is not an acceptable representation of $%
SO(3)$. Spin $\frac{1}{2}$ particles or qubits need to be rotated $720^{0}$
in order to come back to the same state \cite{maggiore}. Topologically, $%
SU(2)$ is the $3$-sphere $\mathcal{S}^{3}$, $SU(2)\approx \mathcal{S}^{3}$.
Instead, $SO(3)$ is topologically equivalent to the projective space $%
\mathbb{R}
P^{3}$ where $%
\mathbb{R}
P^{3}$ results from $\mathcal{S}^{3}$ by identifying pairs of antipodal
points. This leads to conclude the actual isomorphism between groups is $%
SU(2)/%
\mathbb{Z}
_{2}\cong SO(3)$. In formal mathematical terms, there is a not faithful
representation $\varkappa $ of $SU(2)$ as a group of rotations of $%
\mathbb{R}
^{3}$,%
\begin{equation}
\varkappa :SU(2)\ni U_{SU(2)}\left( \vec{A}\text{, }\theta \right) \overset{%
\text{def}}{=}\exp \left( \frac{\vec{\Sigma}}{2i_{%
\mathbb{C}
}}\cdot \vec{A}\theta \right) \mapsto R_{SO(3)}\left( \vec{A}\text{, }\theta
\right) \overset{\text{def}}{=}\exp \left( \vec{E}\cdot \vec{A}\theta
\right) \in SO(3)\text{,}  \label{70}
\end{equation}%
for any vector $\vec{A}=\left( A_{1}\text{, }A_{2}\text{, }A_{3}\right) $.
For the sake of mathematical correctness, we point out that the use of the
dot-notation in (\ref{70})\ (and in the following equations (\ref{72}), (\ref%
{83}), (\ref{jj})) is indeed an abuse of notation for the Euclidean inner
product because $\vec{A}$\ is geometrically just a vector in $%
\mathbb{R}
^{3}$ while $\vec{\Sigma}$ are operators (Pauli matrices) on a
two-dimensional Hilbert space.\textbf{\ }The vector $\vec{E}=\left( E_{1}%
\text{, }E_{2}\text{, }E_{3}\right) $ forms a basis of infinitesimal
generators of the Lie algebra $\mathfrak{so(}3\mathfrak{)}$ of $SO(3)$,%
\begin{equation}
E_{1}=\left( 
\begin{array}{ccc}
0 & 0 & 0 \\ 
0 & 0 & -1 \\ 
0 & 1 & 0%
\end{array}%
\right) \text{, }E_{2}=\left( 
\begin{array}{ccc}
0 & 0 & 1 \\ 
0 & 0 & 0 \\ 
-1 & 0 & 0%
\end{array}%
\right) \text{, }E_{3}=\left( 
\begin{array}{ccc}
0 & -1 & 0 \\ 
1 & 0 & 0 \\ 
0 & 0 & 0%
\end{array}%
\right) \text{,}
\end{equation}%
and they satisfy the following commutation relations, $\left[ E_{l}\text{, }%
E_{m}\right] =\varepsilon _{lmk}E_{k}$. The infinitesimal generators of the
Lie algebra $\mathfrak{su(}2\mathfrak{)}$ of $SU(2)$ are $i_{%
\mathbb{C}
}\vec{\Sigma}=\left( i_{%
\mathbb{C}
}\Sigma _{1}\text{, }i_{%
\mathbb{C}
}\Sigma _{2}\text{, }i_{%
\mathbb{C}
}\Sigma _{3}\right) $ satisfying the commutation relations, $\left[ \Sigma
_{l}\text{, }\Sigma _{m}\right] =2i_{%
\mathbb{C}
}\varepsilon _{lmk}\Sigma _{k}$. This commutations relations are the same as
for $SO(3)$ if one uses $\frac{\Sigma _{l}}{2i_{%
\mathbb{C}
}}$ as the new basis for $\mathfrak{su(}2\mathfrak{)}$. The map $\varkappa $
is exactly $2:1$ and therefore to a rotation of $%
\mathbb{R}
^{3}$ about an axis given by a unit vector $\vec{A}$ through an angle $%
\theta $ radians one associates two $2\times 2$ unitary matrices with
determinant $1$,%
\begin{equation}
\exp \left( \frac{\vec{\Sigma}}{2i_{%
\mathbb{C}
}}\cdot \vec{A}\theta \right) \text{, }\exp \left( \frac{\vec{\Sigma}}{2i_{%
\mathbb{C}
}}\cdot \vec{A}\left( \theta +2\pi \right) \right) \text{.}  \label{72}
\end{equation}%
In other words, $SO(3)$ not only has the usual representation by $3\times 3$
matrices, it also has a double-valued representation by $2\times 2$ matrices
acting on $%
\mathbb{C}
^{2}$. The complex vectors $\left( 
\begin{array}{cc}
\psi ^{1} & \psi ^{2}%
\end{array}%
\right) ^{t}\in 
\mathbb{C}
^{2}$ on which $SO(3)$ acts in this double-valued way are called spinors. In
mathematical terms, $SU(2)$ furnishes naturally a spinor representation of
the $2$-fold cover of $SO(3)$. When $SU(2)$ is thought of as the $2$-fold
cover of $SO(3)$, it is called the spin group $\emph{Spin}(3)$. It is
remarkably powerful to represent three-dimensional rotations in terms of
two-dimensional unitary transformations. In quantum information science,
this is especially true in proving certain circuit identities, in the
characterization of the general $1$-qubit state and in the construction of
the Hardy state \cite{mermin}. Indeed, this representation plays also a key
role in the proof of universality of quantum gates provided by Boykin et 
\textit{al}. For instance, the surjective homeomorphism $\varkappa $ is a
powerful tool for investigating the product of two or more rotations. This
is a consequence of the fact that the Pauli matrices satisfy very simple
product rules, $\Sigma _{l}\Sigma _{m}=\delta _{lm}+i_{%
\mathbb{C}
}\varepsilon _{lmk}\Sigma _{k}$. The infinitesimal generators $\left\{
E_{l}\right\} $ with $l=1$, $2$, $3$ of $\mathfrak{so(}3\mathfrak{)}$ do not
satisfy such simple product relations. For instance, $E_{1}^{2}=diag\left( 0%
\text{, }-1\text{, }-1\right) $.

\subsubsection{Remarks on the rotor group}

One of the most powerful applications of geometric algebra is to rotations.
Within GA, rotations are handled through the use of \emph{rotors}. \ Rotors
also provide a convenient framework for studying Lie groups and Lie
algebras. Let us introduce few definitions. More details on the mathematical
structure of Clifford algebras appears in \cite{porto}

Let $\mathcal{G}\left( p\text{, }q\right) $ denote the GA of a space of
signature $p$, $q$ with $p+q=n$ and let $\mathcal{V}$ be the space of grade-$%
1$ multivectors. Then, the pin group (with respect to the geometric product) 
$\emph{Pin}\left( p\text{, }q\right) $ is defined as,%
\begin{equation}
\emph{Pin}\left( p\text{, }q\right) \overset{\text{def}}{=}\left\{ M\in 
\mathcal{G}\left( p\text{, }q\right) :MaM^{-1}\in \mathcal{V}\text{ }\forall
a\in \mathcal{V}\text{, }MM^{\dagger }=\pm 1\right\} \text{,}
\end{equation}%
where "$\dagger $" denotes the reversion operation in GA (for instance, $%
\left( a_{1}a_{2}\right) ^{\dagger }=a_{2}a_{1}$). The elements of the pin
group split into even-grade and odd-grade elements. The even-grade
multivectors $\left\{ S\right\} $ of the pin group form a subgroup called
the spin group $\emph{Spin}\left( p\text{, }q\right) $,%
\begin{equation}
\emph{Spin}\left( p\text{, }q\right) \overset{\text{def}}{=}\left\{ S\in 
\mathcal{G}_{+}\left( p\text{, }q\right) :SaS^{-1}\in \mathcal{V}\text{ }%
\forall a\in \mathcal{V}\text{, }SS^{\dagger }=\pm 1\right\} \text{,}
\end{equation}%
where $\mathcal{G}_{+}\left( p\text{, }q\right) $ denotes the even
subalgebra of $\mathcal{G}\left( p\text{, }q\right) $. Finally, rotors are
elements $\left\{ R\right\} $ of the spin group satisfying the further
constraint that $RR^{\dagger }=+1$. These elements define the so-called
rotor group $\emph{Spin}^{+}\left( p\text{, }q\right) $,%
\begin{equation}
\emph{Spin}^{+}\left( p\text{, }q\right) \overset{\text{def}}{=}\left\{ R\in 
\mathcal{G}_{+}\left( p\text{, }q\right) :RaR^{\dagger }\in \mathcal{V}\text{
}\forall a\in \mathcal{V}\text{, }RR^{\dagger }=+1\right\} \text{.}
\end{equation}%
For Euclidean spaces, $\emph{Spin}\left( n\text{, }0\right) =\emph{Spin}%
^{+}\left( n\text{, }0\right) $. Therefore, for such spaces, there is no
distinction between the spin group and the rotor group.

In GA, the rotation of a vector $a$ through $\theta $ in the plane generated
by two unit vectors $m$ and $n$ is defined by the double-sided half-angle
transformation law,%
\begin{equation}
a\rightarrow a^{\prime }\overset{\text{def}}{=}RaR^{\dagger }\text{.}
\end{equation}%
The rotor $R$ is defined by,%
\begin{equation}
R\overset{\text{def}}{=}nm=n\cdot m+n\wedge m\equiv \exp \left( -B\frac{%
\theta }{2}\right) \text{,}
\end{equation}%
where the bivector $B$ is such that,%
\begin{equation}
B=\frac{m\wedge n}{\sin \frac{\theta }{2}}\text{, }B^{2}=-1\text{.}
\end{equation}%
Rotors provide a way of handling rotations that is unique to GA and this is
a consequence of the definition of the geometric product. Notice that rotors
are geometric products of two unit vectors and therefore they are
mixed-grade objects. The rotor $R$ on its own has no significance, which is
to say that no meaning should be attached to the separate scalar and
bivector terms. When $R$ is written as the exponential of the bivector $B$
(all rotors near the origin can be written as the exponential of a bivector
and the exponential of a bivector always returns to a rotor), however, the
bivector has a clear geometric significance, as does the vector formed from $%
RaR^{\dagger }$. This illustrates a central feature of GA, which is that
both geometrically meaningful objects (vectors, planes, etc.) and the
elements (operators) that act on them (in this case, rotors or bivectors)
are contained in the same geometric Clifford algebra. Notice that $R$ and $%
-R $ generate the same rotation, so there is a two-to-one map between rotors
and rotations. Formally, the rotor group provides a double-cover
representation of the rotation group.

The Lie algebra of the rotor group $\emph{Spin}^{+}\left( 3\text{, }0\right) 
$ is defined in terms of the bivector algebra,%
\begin{equation}
\left[ B_{l}\text{, }B_{m}\right] =2B_{l}\times B_{m}=-2\varepsilon
_{lmk}B_{k}\text{,}
\end{equation}%
where "$\times $" denotes the commutator product of two multivectors in GA
and,%
\begin{equation}
B_{1}=\sigma _{2}\sigma _{3}=i\sigma _{1}\text{, }B_{2}=\sigma _{3}\sigma
_{1}=i\sigma _{2}\text{, }B_{3}=\sigma _{1}\sigma _{2}=i\sigma _{3}\text{.}
\end{equation}%
The commutator of a bivector with a second bivector produces a third
bivector. That is, the space of bivectors is closed under the commutator
product. This closed algebra defines the Lie algebra of the associated rotor
group. The group is formed by the act of exponentiation. Furthermore, notice
that the product of bivectors satisfies,%
\begin{equation}
B_{l}B_{m}=-\delta _{lm}-\varepsilon _{lmk}B_{k}\text{.}
\end{equation}%
The symmetric part of this product is a scalar, whereas the antisymmetric
part is a bivector. As a final remark, we point out that the algebra of
bivectors is similar to the algebra of the generators of the quaternions.
Thus, quaternions can be identified with bivectors within the GA approach.
The relations among $SO(3)$, $SU(2)$ and the rotor group are summarized as
follows,%
\begin{equation}
\begin{tabular}{||c||c||c||c||}
\hline\hline
\emph{Lie Groups} & \emph{Lie Algebras} & \emph{Product Rules} & \emph{%
Operator},\emph{\ Vectors} \\ \hline\hline
$SO(3)$ & $\left[ E_{l}\text{, }E_{m}\right] =\varepsilon _{lmk}E_{k}$ & not
useful & Orthogonal transformations, vectors in $%
\mathbb{R}
^{3}$ \\ \hline\hline
$SU(2)$ & $\left[ \Sigma _{l}\text{, }\Sigma _{m}\right] =2i_{%
\mathbb{C}
}\varepsilon _{lmk}\Sigma _{k}$ & $\Sigma _{l}\Sigma _{m}=\delta _{lm}+i_{%
\mathbb{C}
}\varepsilon _{lmk}\Sigma _{k}$ & Unitary operators, spinors \\ \hline\hline
$\emph{Spin}^{+}\left( 3\text{, }0\right) $ & $\left[ B_{l}\text{, }B_{m}%
\right] =-2\varepsilon _{lmk}B_{k}$ & $B_{l}B_{m}=-\delta _{lm}-\varepsilon
_{lmk}B_{k}$ & Rotors (or, bivectors), multivectors \\ \hline\hline
\end{tabular}%
\text{.}
\end{equation}%
Therefore, two central features of GA emerge: 1) the GA provides a very
clear and compact method for encoding rotations which is considerably more
powerful than working with matrices; 2) Both geometrically meaningful
objects (vectors, planes, etc.) and the elements (operators) that act on
them (in this case, rotors or bivectors) are contained in the same geometric
Clifford algebra.

\subsection{Universal Sets of Quantum Gates}

Quantum computational gates are input-output devices whose inputs and
outputs are discrete quantum variables such as spins. As a matter of fact,
recall that the most general $2\times 2$ unitary matrix with determinant $1$
can be expressed in the form of a finite rotation represented as,%
\begin{equation}
\hat{U}_{SU\left( 2\right) }\left( \hat{n}\text{, }\theta \right) \overset{%
\text{def}}{=}e^{-i_{%
\mathbb{C}
}\frac{\theta }{2}\hat{n}\cdot \vec{\Sigma}}=\hat{I}\cos \left( \frac{\theta 
}{2}\right) -i_{%
\mathbb{C}
}\hat{n}\cdot \vec{\Sigma}\sin \left( \frac{\theta }{2}\right) \text{.}
\label{83}
\end{equation}%
Therefore, we are entitled to think of a qubit as the state of a spin-$\frac{%
1}{2}$ object and an arbitrary unitary transformation (quantum gate) acting
on the state (aside from a possible rotation of the overall phase) is a
rotation of the spin. A set of gates is \emph{adequate} if any quantum
computation can be performed with arbitrary precision by networks consisting
only of replicas of gates from that set. A gate is \emph{universal} if by
itself it forms an adequate set, i. e. if any quantum computation can be
performed by a network containing replicas of only this gate. The first
example of universal gate is the universal Deutsch three-bit gate \cite{D89}%
. In the network's computational basis $\mathcal{B}_{\mathcal{H}%
_{2}^{3}}=\left\{ \left\vert 000\right\rangle \text{,}\left\vert
100\right\rangle \text{, }\left\vert 010\right\rangle \text{, }\left\vert
001\right\rangle \text{, }\left\vert 110\right\rangle \text{, }\left\vert
101\right\rangle \text{, }\left\vert 011\right\rangle \text{, }\left\vert
111\right\rangle \right\} $, it is given by the $8\times 8$ unitary matrix $%
\mathcal{D}_{\text{universal}}^{\left( \text{Deutsch}\right) }\left( \gamma
\right) $,%
\begin{equation}
\mathcal{D}_{\text{universal}}^{\left( \text{Deutsch}\right) }\left( \gamma
\right) \overset{\text{def}}{=}\left( 
\begin{array}{cc}
I_{6\times 6} & O_{6\times 2} \\ 
O_{2\times 6} & D_{2\times 2}\left( \gamma \right)%
\end{array}%
\right) \text{,}
\end{equation}%
where $I_{l\times k}$ is the $l\times k$ identity matrix, $O_{l\times k}$ is
the $l\times k$ null matrix and $D_{2\times 2}\left( \gamma \right) $ is
defined as,%
\begin{equation}
D_{2\times 2}\left( \gamma \right) \overset{\text{def}}{=}\left( 
\begin{array}{cc}
i_{%
\mathbb{C}
}\cos \left( \frac{\pi \gamma }{2}\right) & \sin \left( \frac{\pi \gamma }{2}%
\right) \\ 
\sin \left( \frac{\pi \gamma }{2}\right) & i_{%
\mathbb{C}
}\cos \left( \frac{\pi \gamma }{2}\right)%
\end{array}%
\right) \text{.}
\end{equation}%
The Deutsch gate depends on the parameter $\gamma $ that can be any
irrational number. Another important example of universal quantum gate is
given by the Barenco three-parameter family of universal two-bit gates \cite%
{B95}. In the network's computational basis $\mathcal{B}_{\mathcal{H}%
_{2}^{2}}=\left\{ \left\vert 00\right\rangle \text{,}\left\vert
10\right\rangle \text{, }\left\vert 01\right\rangle \text{, }\left\vert
11\right\rangle \right\} $, it is given by the $4\times 4$ unitary matrix $%
\mathcal{A}_{\text{universal}}^{\left( \text{Barenco}\right) }\left( \phi 
\text{, }\alpha \text{, }\theta \right) $,%
\begin{equation}
\mathcal{A}_{\text{universal}}^{\left( \text{Barenco}\right) }\left( \phi 
\text{, }\alpha \text{, }\theta \right) \overset{\text{def}}{=}\left( 
\begin{array}{cc}
I_{2\times 2} & O_{2\times 2} \\ 
O_{2\times 2} & A_{2\times 2}\left( \phi \text{, }\alpha \text{, }\theta
\right)%
\end{array}%
\right) \text{,}
\end{equation}%
where $I_{l\times k}$ is the $l\times k$ identity matrix, $O_{l\times k}$ is
the $l\times k$ null matrix and $A_{2\times 2}\left( \phi \text{, }\alpha 
\text{, }\theta \right) $ is defined as, 
\begin{equation}
A_{2\times 2}\left( \phi \text{, }\alpha \text{, }\theta \right) \overset{%
\text{def}}{=}\left( 
\begin{array}{cc}
e^{i_{%
\mathbb{C}
}\alpha }\cos \theta & -i_{%
\mathbb{C}
}e^{i_{%
\mathbb{C}
}\left( \alpha -\phi \right) }\sin \theta \\ 
-i_{%
\mathbb{C}
}e^{i_{%
\mathbb{C}
}\left( \alpha +\phi \right) }\sin \theta & e^{i_{%
\mathbb{C}
}\alpha }\cos \theta%
\end{array}%
\right) \text{.}
\end{equation}%
The Barenco gate depends on three parameters $\phi $, $\alpha $, and $\theta 
$ that are fixed irrational multiples of $\pi $ and of each other. More
generally, it turns out that \emph{almost }all\emph{\ }two-bit (and more
inputs) quantum gates are universal \cite{D95, L96}.

A set of quantum gates $\mathcal{S}$ is said to be universal if an arbitrary
unitary quantum operation can be performed with arbitrarily small error
probability using a quantum circuit that only uses gates from $\mathcal{S}$.
An important set of logic gates in quantum computing is given by,%
\begin{equation}
\mathcal{S}_{\text{Clifford}}\overset{\text{def}}{=}\left\{ \hat{H}\text{, }%
\hat{P}\text{, }\hat{U}_{\text{CNOT}}\right\} \text{.}
\end{equation}%
The set $\mathcal{S}_{\text{Clifford}}$ of Hadamard-$\hat{H}$, phase-$\hat{P}
$ and CNOT-$U_{\text{CNOT}}$ gates generates the so-called Clifford group,
the normalizer $\mathcal{N}\left( \mathcal{G}_{n}\right) $ of the Pauli
group $\mathcal{G}_{n}$ in $\mathcal{U}\left( n\right) $ \cite{C98}. This
set of gates is sufficient to perform fault-tolerant quantum computation but 
$\mathcal{S}_{\text{Clifford}}$ is not sufficiently powerful to perform
universal quantum computation. However, universal quantum computation
becomes possible if the gates in the Clifford group are supplemented with
the Toffoli gate \cite{S96},%
\begin{equation}
\mathcal{S}_{\text{universal}}^{\left( \text{Shor}\right) }\overset{\text{def%
}}{=}\left\{ \hat{H}\text{, }\hat{P}\text{, }\hat{U}_{\text{CNOT}}\text{, }%
\hat{U}_{\text{Toffoli}}\right\} \text{.}
\end{equation}%
Shor showed that adding the Toffoli gate to the generators of the Clifford
group produces the universal set $\mathcal{S}_{\text{universal}}^{\left( 
\text{Shor}\right) }$. Another example of universal set of logic gates is
provided by Boykin et \textit{al.} in \cite{B99, B00}. The set they
construct is given by,%
\begin{equation}
\mathcal{S}_{\text{universal}}^{\left( \text{Boykin et \textit{al}.}\right) }%
\overset{\text{def}}{=}\left\{ \hat{H}\text{, }\hat{P}\text{, }\hat{T}\text{%
, }\hat{U}_{\text{CNOT}}\text{ }\right\} \text{.}
\end{equation}%
This set is presumably easier to implement experimentally than $\mathcal{S}_{%
\text{universal}}^{\left( \text{Shor}\right) }$ since the $\hat{T}$ is a
one-qubit gate while the Toffoli gate is a three-qubit gate.

\subsection{GA reexamination of Boykin's proof of universality}

Boykin's proof is very elegant and is solely based on the geometry of real
rotations in three dimensions and on the local isomorphism between the Lie
groups $SO(3)$ and $SU(2)$. In what follows, we will revisit the proof using
a GA approach based on the rotor group $\emph{Spin}^{+}\left( 3\text{, }%
0\right) $ and on the algebra of bivectors, $\left[ B_{l}\text{, }B_{m}%
\right] =-2\varepsilon _{lmk}B_{k}$.

The proof of universality of the basis $\mathcal{S}_{\text{universal}%
}^{\left( \text{Boykin et \textit{al}.}\right) }$ can be presented in two
steps. In the first step, it is required to show that Hadamard gate $\hat{H}$
and the $\frac{\pi }{8}$-phase gate $\hat{T}=$ $\hat{\Sigma}_{3}^{\frac{1}{4}%
}$ form a \emph{dense} set in $SU(2)$ where, 
\begin{equation}
\hat{\Sigma}_{3}^{\alpha }\overset{\text{def}}{=}\left( 
\begin{array}{cc}
1 & 0 \\ 
0 & e^{i_{%
\mathbb{C}
}\pi \alpha }%
\end{array}%
\right) \text{, }\hat{\Sigma}_{3}^{\alpha }\left\vert \psi \right\rangle
\leftrightarrow \psi _{\hat{\Sigma}_{3}^{\alpha }}^{\left( \text{GA}\right)
}=\sigma _{3}^{\alpha }\psi \sigma _{3}\text{. }
\end{equation}%
This means that any element $\hat{U}_{SU(2)}$ in $SU(2)$ can be approximated
to a desired degree of precision by a finite product of $\hat{H}$ and $\hat{T%
}$. In other words, when a circuit of quantum gates is used to implement
some desired unitary operation $\hat{U}$, it is sufficient to have an
implementation that approximates $\hat{U}$ to some specified level of
accuracy. Suppose we approximate $\hat{U}$ by some other unitary
transformation $\hat{U}^{\prime }$. Then, the notion of the quality of an
approximation of a unitary transformation can be quantified considering the
so-called approximation error $\varepsilon \left( \hat{U}\text{, }\hat{U}%
^{\prime }\right) $ \cite{afamm},%
\begin{equation}
\varepsilon \left( \hat{U}\text{, }\hat{U}^{\prime }\right) \overset{\text{%
def}}{=}\max_{\left\vert \psi \right\rangle }\left\Vert \left( \hat{U}-\hat{U%
}^{\prime }\right) \left\vert \psi \right\rangle \right\Vert \text{,}
\end{equation}%
where $\left\Vert \psi \right\Vert =\sqrt{\left\langle \psi |\psi
\right\rangle }$ is the Euclidean norm of $\left\vert \psi \right\rangle $
and $\left\langle \cdot |\cdot \right\rangle $ is the conventional inner
product defined on the complex Hilbert space. In the second step of the
proof, it is necessary to point out that for universal computation all that
is needed is $\hat{U}_{\text{CNOT}}$ and $SU(2)$ \cite{barenco}.

To show that $\hat{H}$ and $\hat{T}$ form a \emph{dense} set in $SU(2)$, the
local isomorphism between $SO(3)$ and $SU(2)$ must be exploited. Indeed, it
can be shown that using the set $\left\{ \hat{H}\text{, }\hat{T}=\hat{\Sigma}%
_{3}^{\frac{1}{4}}\right\} $, we can construct quantities in this basis that
correspond to rotations by angles that are irrational multiples of $\pi $ in 
$SO(3$, $%
\mathbb{R}
)$ about two orthogonal axes. Consider the following two rotations in $SO(3)$
described in terms of rotors in $\emph{Spin}^{+}\left( 3\text{, }0\right) $,%
\begin{equation}
SO(3)\ni \hat{U}_{SO(3)}^{\left( 1\right) }\overset{\text{def}}{=}e^{i_{%
\mathbb{C}
}\lambda _{1}\pi \hat{n}_{1}\cdot \vec{\Sigma}}\leftrightarrow
e^{in_{1}\lambda _{1}\pi }\in \emph{Spin}^{+}\left( 3\text{, }0\right) \text{%
, }\hat{U}_{SO(3)}^{\left( 2\right) }\overset{\text{def}}{=}e^{i_{%
\mathbb{C}
}\lambda _{2}\pi \hat{n}_{2}\cdot \vec{\Sigma}}\leftrightarrow
e^{in_{2}\lambda _{2}\pi }\text{, }  \label{111}
\end{equation}%
where the elements $\lambda _{1}$, $\lambda _{2}$ are irrational numbers in $%
\mathbb{R}
/%
\mathbb{Q}
$. Let us show that rotations in (\ref{111}) can be expressed in terms of a
suitable combination of elements in $\left\{ \hat{H}\text{, }\hat{T}=\hat{%
\Sigma}_{3}^{\frac{1}{4}}\right\} $. It turns out that since $SU(2)/%
\mathbb{Z}
_{2}\cong SO(3)$, we have 
\begin{equation}
\emph{Spin}^{+}\left( 3\text{, }0\right) \ni e^{in_{1}\lambda _{1}\pi
}\leftrightarrow \hat{U}_{SU(2)}^{\left( 1\right) }\overset{\text{def}}{=}%
\hat{\Sigma}_{3}^{-\frac{1}{4}}\hat{\Sigma}_{1}^{\frac{1}{4}}\in SU(2)\text{
and, }e^{in_{2}\lambda _{2}\pi }\leftrightarrow \hat{U}_{SU(2)}^{\left(
2\right) }\overset{\text{def}}{=}\hat{H}^{-\frac{1}{2}}\hat{\Sigma}_{3}^{-%
\frac{1}{4}}\hat{\Sigma}_{1}^{\frac{1}{4}}\hat{H}^{\frac{1}{2}}\text{,}
\label{m}
\end{equation}%
where $\hat{\Sigma}_{1}^{\frac{1}{4}}=\hat{H}\hat{\Sigma}_{3}^{\frac{1}{4}}%
\hat{H}$. Working out the details of \cite{B99, B00} and using the results
presented in Section III, it turns out that the rotor representation of $%
\hat{U}_{SU(2)}^{\left( 1\right) }$ and $\hat{U}_{SU(2)}^{\left( 2\right) }$
is given by,%
\begin{equation}
\hat{U}_{SU(2)}^{\left( 1\right) }\leftrightarrow R_{1}=\frac{1}{2}\left( 1+%
\frac{1}{\sqrt{2}}\right) \mathbf{-}\frac{1}{2\sqrt{2}}i\sigma _{1}+\frac{1}{%
2}\left( 1-\frac{1}{\sqrt{2}}\right) i\sigma _{2}+\frac{1}{2\sqrt{2}}i\sigma
_{3}\text{,}
\end{equation}%
and,%
\begin{equation}
\hat{U}_{SU(2)}^{\left( 2\right) }\leftrightarrow R_{2}=\frac{1}{2}\left( 1+%
\frac{1}{\sqrt{2}}\right) \mathbf{-}\frac{1}{2}\left( \frac{1}{2}-\frac{1}{%
\sqrt{2}}\right) i\sigma _{1}+\frac{1}{2}i\sigma _{2}+\frac{1}{2}\left( 
\frac{1}{2}-\frac{1}{\sqrt{2}}\right) i\sigma _{3}\text{,}
\end{equation}%
respectively. $R_{1}$ and $R_{2}$ are rotors in $\emph{Spin}^{+}\left( 3%
\text{, }0\right) $. Notice that,%
\begin{equation}
e^{in_{k}\lambda _{k}\pi }=\cos \left( \lambda _{k}\pi \right) +n_{kx}\sin
\left( \lambda _{k}\pi \right) i\sigma _{1}+n_{ky}\sin \left( \lambda
_{k}\pi \right) i\sigma _{2}+n_{kz}\sin \left( \lambda _{k}\pi \right)
i\sigma _{3}\text{,}
\end{equation}%
for $k=1$, $2$ and unit vectors $n_{k}$. Therefore, setting $%
e^{in_{1}\lambda _{1}\pi }=R_{1}$ we get,%
\begin{equation}
\cos \left( \lambda _{1}\pi \right) =\frac{1}{2}\left( 1+\frac{1}{\sqrt{2}}%
\right) \text{, }n_{1y}\sin \left( \lambda _{1}\pi \right) =\frac{1}{2}%
\left( 1-\frac{1}{\sqrt{2}}\right) \text{, }n_{1z}\sin \left( \lambda
_{1}\pi \right) =\frac{1}{2}\frac{1}{\sqrt{2}}\text{, }n_{1x}=-n_{1z}\text{.}
\end{equation}%
Finally, after some algebra, we obtain that $\lambda _{1}$ is equal to,%
\begin{equation}
\lambda _{1}=\frac{1}{\pi }\cos ^{-1}\left[ \frac{1}{2}\left( 1+\frac{1}{%
\sqrt{2}}\right) \right] \text{,}
\end{equation}%
and the unit vector $n_{1}=n_{1x}\sigma _{1}+n_{1y}\sigma _{2}+n_{1z}\sigma
_{3}$ is such that,%
\begin{equation}
\left( n_{1x}\text{, }n_{1y}\text{, }n_{1z}\right) =\frac{1}{\sqrt{1-\left[ 
\frac{1}{2}\left( 1+\frac{1}{\sqrt{2}}\right) \right] ^{2}}}\left( -\frac{1}{%
2\sqrt{2}}\text{, }\frac{1}{2}(1-\frac{1}{\sqrt{2}})\text{, }\frac{1}{2\sqrt{%
2}}\text{ }\right) \text{.}  \label{BB}
\end{equation}%
Similarly, setting $e^{in_{1}\lambda _{2}\pi }=R_{2}$, we obtain,%
\begin{equation}
\cos \left( \lambda _{2}\pi \right) =\frac{1}{2}\left( 1+\frac{1}{\sqrt{2}}%
\right) \text{, }n_{2y}\sin \left( \lambda _{2}\pi \right) =\frac{1}{2}\text{%
, }n_{2z}\sin \left( \lambda _{2}\pi \right) =\frac{1}{2}\left( \frac{1}{2}-%
\frac{1}{\sqrt{2}}\right) \text{, }n_{2x}=-n_{2z}\text{.}
\end{equation}%
Finally, after some algebra, we obtain that $\lambda _{2}=\lambda _{1}$ is
equal to,%
\begin{equation}
\lambda _{2}=\frac{1}{\pi }\cos ^{-1}\left[ \frac{1}{2}\left( 1+\frac{1}{%
\sqrt{2}}\right) \right] \text{,}
\end{equation}%
and the unit vector $n_{2}=n_{2x}\sigma _{1}+n_{2y}\sigma _{2}+n_{2z}\sigma
_{3}$ is such that,%
\begin{equation}
\left( n_{2x}\text{, }n_{2y}\text{, }n_{2z}\right) =\frac{1}{\sqrt{1-\left[ 
\frac{1}{2}\left( 1+\frac{1}{\sqrt{2}}\right) \right] ^{2}}}\left( -\frac{1}{%
2}(\frac{1}{2}-\frac{1}{\sqrt{2}})\text{, }\frac{1}{2}\text{, }\frac{1}{2}(%
\frac{1}{2}-\frac{1}{\sqrt{2}})\text{ }\right) \text{.}  \label{BBB}
\end{equation}%
From (\ref{BB}) and (\ref{BBB}), it follows that $n_{1}\cdot n_{2}=0$. Since 
$\lambda _{1}=\lambda _{2}\equiv \lambda \in 
\mathbb{R}
/%
\mathbb{Q}
$, there exist some $n\in 
\mathbb{N}
$ such that any phase factor $e^{i_{%
\mathbb{C}
}\phi }$ can be approximated by $e^{i_{%
\mathbb{C}
}n\lambda \pi }$,%
\begin{equation}
e^{i_{%
\mathbb{C}
}\phi }\approx e^{i_{%
\mathbb{C}
}n\lambda \pi }\text{, }n\in 
\mathbb{N}
\text{.}  \label{mmm}
\end{equation}%
From (\ref{m}) and (\ref{mmm}), it follows that we have at least two dense
subsets of $SU(2$, $%
\mathbb{C}
)$, that is to say $e^{in_{1}\alpha }$ and $e^{i\beta n_{2}}$ with,%
\begin{equation}
\alpha \approx \lambda \pi l\text{ (mod}2\pi \text{) and, }\beta \approx
\lambda \pi l\text{ (mod}2\pi \text{) with }l\in 
\mathbb{N}
\text{.}
\end{equation}%
Since $n_{1}$ and $n_{2}$ are orthogonal vectors, we can write any element $%
\hat{U}_{SU(2)}$ $\in $ $SU(2$, $%
\mathbb{C}
)$ in the following form,%
\begin{equation}
\hat{U}_{SU(2)}=e^{i_{%
\mathbb{C}
}\phi \hat{n}\cdot \vec{\Sigma}}\leftrightarrow e^{in\phi }=e^{in_{1}\alpha
}e^{in_{2}\beta }e^{in_{1}\gamma }\text{.}  \label{jj}
\end{equation}%
Notice that the representation in (\ref{jj}) is analogous to Euler rotations
about three orthogonal vectors. Expansion of the LHS of (\ref{jj}) leads to,%
\begin{equation}
e^{in\phi }=\cos \phi +in\sin \phi \text{.}  \label{cx}
\end{equation}%
Expansion of the RHS of (\ref{jj}) yields,%
\begin{equation}
e^{in_{1}\alpha }e^{in_{2}\beta }e^{in_{1}\gamma }=\left( \cos \alpha
+in_{1}\sin \alpha \right) \left( \cos \beta +in_{2}\sin \beta \right)
\left( \cos \gamma +in_{1}\sin \gamma \right) \text{.}  \label{yes}
\end{equation}%
Recalling that $n_{1}n_{2}=n_{1}\cdot n_{2}+n_{1}\wedge n_{2}$ and that the
unit vectors $n_{1}$ and $n_{2}$ are orthogonal, we have,%
\begin{equation}
n_{1}n_{2}=-n_{2}n_{1}\text{.}  \label{too}
\end{equation}%
Moreover, recalling that,%
\begin{equation}
\sin \left( \alpha \pm \beta \right) =\sin \alpha \cos \beta \pm \cos \alpha
\sin \beta \text{, and, }\cos \left( \alpha \pm \beta \right) =\cos \alpha
\cos \beta \mp \sin \alpha \sin \beta \text{, }  \label{xxx}
\end{equation}%
further expansion of (\ref{yes}) together with the use of (\ref{too})\ and (%
\ref{xxx}), leads to%
\begin{equation}
e^{in\phi }=\cos \beta \cos \left( \alpha +\gamma \right) +\cos \beta \sin
\left( \alpha +\gamma \right) in_{1}+\sin \beta \cos \left( \gamma -\alpha
\right) in_{2}+\sin \beta \sin \left( \gamma -\alpha \right) n_{1}\wedge
n_{2}\text{.}  \label{cxx}
\end{equation}%
Setting (\ref{cx}) equal to (\ref{cxx}), we finally obtain%
\begin{equation}
\cos \phi =\cos \beta \cos \left( \alpha +\gamma \right)  \label{A}
\end{equation}%
and,%
\begin{equation}
n\sin \phi =\cos \beta \sin \left( \alpha +\gamma \right) n_{1}+\sin \beta
\cos \left( \gamma -\alpha \right) n_{2}-i\sin \beta \sin \left( \gamma
-\alpha \right) \left( n_{1}\wedge n_{2}\right) \text{.}  \label{B}
\end{equation}%
In conclusion, the parameters $\alpha $, $\beta $ and $\gamma $ can be found
by inverting (\ref{A}) and (\ref{B}) for any element in $SU(2)$. Then, using
the fact that $\hat{U}_{\text{CNOT}}$ and $SU(2)$ form a universal basis for
quantum computing \cite{barenco}, the proof is completed \cite{B99, B00}. It
is evident that the GA provides a very clear and compact method for encoding
rotations which is considerably more powerful than working with matrices.
Furthermore, from a conceptual point of view, a central feature of GA
emerges as well\textbf{\ (}although such feature appears in most geometric
algebra applications\textbf{)}: both vectors (grade-$1$ multivectors),
planes (grade-$2$ multivectors) and the operators acting on them (in this
case, rotors $R$ or bivectors $B$) are contained in the same geometric
Clifford algebra.

\section{Conclusions and Remarks}

In this article, we investigated the utility of GA methods in two specific
applications to quantum information science. First, we presented an explicit
multiparticle spacetime algebra description of one and two-qubit quantum
states together with a MSTA characterization of one and two-qubit quantum
computational gates. Second, using the above mentioned explicit
characterization and the GA description of the Lie algebras $SO\left(
3\right) $ and $SU\left( 2\right) $ based on the rotor group $\emph{Spin}%
^{+}\left( 3\text{, }0\right) $ formalism, we reexamined Boykin's proof of
universality of quantum gates. We conclude that the MSTA approach leads to a
useful conceptual unification where the complex qubit space and the complex
space of unitary operators acting on them become united, with both being
made just by multivectors in real space \cite{laser}. Furthermore, the GA
approach to rotations based on the rotor group clearly brings conceptual and
computational advantages compared to standard vectorial and matricial
approaches.\textbf{\ }In what follows, we present few concluding remarks.

In standard quantum computation, the basic operation is the tensor product%
\emph{\ }"$\otimes $". In the GA\ approach to quantum computing, the basic
operation becomes the geometric (Clifford) product. Tensor product has no
neat geometric visualization while geometric product has clear geometric
interpretations. For instance, it forms a cube $\left( \sigma _{1}\sigma
_{2}\sigma _{3}\right) $ from a vector $\left( \sigma _{1}\right) $ and a
square $\left( \sigma _{2}\sigma _{3}\right) $, an oriented square $\left(
\sigma _{1}\sigma _{2}\right) $ from two vectors $\left( \sigma _{1}\text{
and }\sigma _{2}\right) $, a square $\left( \sigma _{2}\sigma _{3}\right) $
from a cube $\left( \sigma _{1}\sigma _{2}\sigma _{3}\right) $ and a vector $%
\left( \sigma _{1}\right) $, and so on. Furthermore, entangled quantum
states are replaced by multivectors with a clear geometric interpretation.
For instance, a general multivector $M$ in $\mathfrak{Cl}(3)$ is a linear
combination of blades, geometric products of different basis vectors
supplemented by the identity $1$ (basic oriented scalar),%
\begin{equation}
M\overset{\text{def}}{=}M_{0}1+\sum_{j=1}^{3}M_{j}\sigma
_{j}+\sum_{j<k}M_{jk}\sigma _{j}\sigma _{k}+M_{123}\sigma _{1}\sigma
_{2}\sigma _{3}\text{, with }j\text{, }k=1\text{, }2\text{, }3\text{.}
\end{equation}%
Entangled states are replaced by GA multivectors that are nothing but bags
of shapes (points, $1$; lines, $\sigma _{j}$; squares, $\sigma _{j}\sigma
_{k}$; cubes, $\sigma _{1}\sigma _{2}\sigma _{3}$; and so on).

As a final remark, we point out that one of the most fascinating open
problems in quantum physics is the characterization of the complexity of
quantum motion. In quantum information science, the concept of complexity is
also defined for quantum unitary operators, the so-called quantum gate
complexity (a quantitative measure for the computational work needed to
accomplish a given task \cite{NIELSEN}). We believe that the conceptual
unification between spaces of \ quantum states and of quantum unitary
operators acting on such states provided by multiparticle geometric algebras
may allow also for the possibility of providing a single mathematical
framework where complexities of both quantum states and quantum gates are
defined for quantities both belonging to the same \emph{real} multivectorial
space (the reality of the multivectorial space is required for geometric
purposes). Furthermore, this unification may turn out to be very useful in
view of the recent connection made between quantum gate complexity and
complexity of the motion on a suitable Riemannian manifold of multi-qubit
unitary transformations provided by Nielsen and coworkers \cite{nc, nc1}.

In view of such considerations, we believe that the application of geometric
Clifford algebras to the characterization of quantum gate complexity and to
quantum information science in general is worthy of further investigations.

\begin{acknowledgments}
C. C. thanks C. Lupo and L. Memarzadeh for very useful discussions. This
work was supported by the European Community's Seventh Framework Program
(CORNER Project; FP7/2007-2013) under grant agreement 213681.
\end{acknowledgments}


\begin{thebibliography}{99}
\bibitem{hestenes} D. Hestenes, "\emph{Spacetime Algebra}", Gordon and
Breach, New York (1966).

\bibitem{dl} C. Doran and A. Lasenby, "\emph{Geometric Algebra for Physicists%
}", Cambridge University Press (2003).

\bibitem{gull} A. Lasenby, C. Doran and S. Gull, "\emph{Gravity, Gauge
Theories and Geometric Algebra}", Phil. Trans. Roy. Soc. Lond. \textbf{A356}%
, 487 (1998).

\bibitem{francis} M. R. Francis and A. Kosowsky, "\emph{Geometric algebra
techniques for general relativity}", Annals of Physics \textbf{311}, 459
(2004).

\bibitem{baylis} W. E. Baylis, "\emph{Electrodynamics: A Modern Geometric
Approach}", World Scientific (1988).

\bibitem{cafaro-ali} C. Cafaro and S. A. Ali, "\emph{The Spacetime Algebra
Approach to Massive Classical Electrodynamics with Magnetic Monopoles}",
Adv.\ Appl. Clifford Alg. \textbf{17}, 23 (2007).

\bibitem{cafaro} C. Cafaro, "\emph{Finite-Range Electromagnetic Interaction
and Magnetic Charges: Spacetime Algebra or Algebra of Physical Space?}",
Adv.\ Appl. Clifford Alg. \textbf{17}, 617 (2007).

\bibitem{vlasov} A. Yu. Vlasov, "\emph{Quantum Gates and Clifford Algebras}%
", arXiv:quant-ph/9907079 (1999).

\bibitem{doran1} T. F. Havel and C. Doran, "\emph{Geometric Algebra in
Quantum Information Processing}", quant-ph/0004031 (2001).

\bibitem{somaroo} S. S. Somaroo et \textit{al}., "\emph{Expressing the
operations of quantum computing in multiparticle geometric algebra}", Phys.
Lett. \textbf{A240}, 1 (1998).

\bibitem{marek1} D. Aerts and M. Czachor, "\emph{Cartoon computation:
quantum-like computing without quantum mechanics}", J. Phys. \textbf{A40},
259 (2007).

\bibitem{marek2} M. Czachor, "\emph{Elementary gates for cartoon computation}%
", J. Phys. \textbf{A40}, 753 (2007).

\bibitem{marek3} D. Aerts and M. Czachor, "\emph{Tensor-product versus
geometric-product coding}", Phys. Rev. \textbf{A77}, 012316 (2008).

\bibitem{lasenby93} A. Lasenby et \textit{al}., "\emph{2-Spinors, Twistors
and Supersymmetry in the Spacetime Algebra}", in Z. Oziewicz et \textit{al}%
., eds., "\emph{Spinors, Twistors, Clifford Algebras and Quantum Deformations%
}" (Kluwer Academic, Dordrecht, 1993), p. 233.

\bibitem{cd93} C. Doran et \textit{al}., "\emph{States and Operators in the
Spacetime Algebra}", Found. Phys. \textbf{23}, 1239 (1993).\textbf{\ }

\bibitem{doran96} C. Doran et \textit{al}., "\emph{Spacetime algebra and
electron physics}", Adv. Imaging Electron Phys. \textbf{95}, 271 (1996).

\bibitem{somaroo99} S. Somaroo et \textit{al}., "\emph{Geometric algebra and
the causal approach to multiparticle quantum mechanics}", J. Math. Phys. 
\textbf{40}, 3327 (1999).

\bibitem{NIELSEN} A. Nielsen and I. L. Chuang, "\emph{Quantum Computation
and Information}", Cambridge University Press (2000).

\bibitem{B99} P. O. Boykin et \textit{al}., "\emph{On universal and
fault-tolerant quantum computing}", in Proceedings of the \textit{40th}
Annual Symposium on Fundamentals of Computer Science, IEEE Press, Los
Alamitos-CA (1999).

\bibitem{B00} P. O. Boykin et \textit{al}., "\emph{A new universal and
fault-tolerant quantum basis}", Information Processing Letters \textbf{75},
101 (2000).

\bibitem{david75} D. Hestenes, "\emph{Observables, Operators, and Complex
Numbers in the Dirac Theory}", J. Math. Phys. \textbf{16}, 556 (1975).

\bibitem{holland} P. R. Holland,\emph{\ "Causal interpretation of a system
of two spin-}$\frac{1}{2}$\emph{\ particles}", Phys. Rep. \textbf{169}, 294
(1988).

\bibitem{morton} M. Hamermesh, "\emph{Group Theory and Its Applications to
Physical Problems}", Dover Publication, Inc., New York (1968).

\bibitem{maggiore} M.\ Maggiore, "\emph{A Modern Introduction to Quantum
Field Theory}", Oxford University Press (2005).

\bibitem{mermin} N. D. Mermin, "\emph{Quantum Computer Science}", Cambridge
University Press (2007).

\bibitem{porto} I. Porteous, "\emph{Mathematical Structure of Clifford
Algebras}", in "Lectures on Clifford (Geometric) Algebras and Applications",
R. Ablamowicz and G. Sobczyk, Editors; Birkh\"{a}user (2004).

\bibitem{D89} D. Deutsch, "\emph{Quantum computational networks}", Proc. R.
Soc. Lond. \textbf{A425}, 73 (1989).

\bibitem{B95} A. Barenco, "\emph{A Universal Two-Bit Gate for Quantum
Computation}", Proc. R. Soc. Lond. \textbf{A449}, 679 (1995).

\bibitem{D95} D. Deutsch et \textit{al}., "\emph{Universality in quantum
computation}", Proc. R. Soc. Lond. \textbf{A449}, 669 (1995).

\bibitem{L96} S. Lloyd, "\emph{Almost any quantum logic gate is universal}",
Phys. Rev. Lett. \textbf{75}, 346 (1996).

\bibitem{C98} R. Calderbank et \textit{al}., "\emph{Quantum error correction
via codes over }$GF\left( 4\right) $", IEEE Trans. Inf. Theor. \textbf{44},
1369 (1998).

\bibitem{S96} P.\ Shor, "\emph{Fault-tolerant quantum computation}", in
Proceedings of the \textit{37th} Annual Symposium on Fundamentals of
Computer Science, IEEE Press, Los Alamitos-CA (1996).

\bibitem{afamm} P. Kaye, R. Laflamme and M. Mosca, "\emph{An Introduction to
Quantum Computing}", Oxford University Press (2007).

\bibitem{barenco} A. Barenco et \textit{al}., "\emph{Elementary gates for
quantum computation}", Phys. Rev. \textbf{A52}, 3457 (1995).

\bibitem{laser} A. Lasenby et \textit{al}., "\emph{STA and the
Interpretation of Quantum Mechanics}", in "Clifford (Geometric) Algebras
with Applications in Physics, Mathematics and Engineering", W. E. Baylis,
Editor; Birkh\"{a}user (1996).

\bibitem{nc} M. A. Nielsen et. \textit{al}., "Quantum Computation as
Geometry", Science \textbf{311}, 1133 (2006).

\bibitem{nc1} M. R. Dowling and M. A. Nielsen, "\emph{The Geometry of
Quantum Computation}", Quantum Information \& Computation \textbf{8}, 0861
(2008).
\end{thebibliography}
\end{document}